\documentclass[sigconf]{acmart}
\usepackage{xcolor}
\usepackage{algorithm}
\usepackage{algpseudocode}
\usepackage{multirow}
\usepackage[table]{xcolor}
\definecolor{myblue}{RGB}{231,239,250}
\definecolor{myyellow}{RGB}{255,226,206}
\definecolor{mygreen}{RGB}{219,233,227}
\definecolor{mypink}{RGB}{252,232,230}
\usepackage{placeins}
\usepackage{graphicx}
\usepackage{subcaption}
\usepackage{dblfloatfix} 
\usepackage{threeparttable}
\usepackage{enumitem}
\usepackage{booktabs}
\usepackage[table]{xcolor}
\usepackage{array}
\usepackage{multirow}

\usepackage{array}
\newcolumntype{C}[1]{>{\centering\arraybackslash}p{#1}}

\renewcommand\footnotetextcopyrightpermission[1]{}
\begin{document}
\pagestyle{plain}

\title{Detecting and Mitigating Backdoor Attacks in OTA-FL \\ Systems: A Two-Stage Robust Aggregation Scheme}

\author{Xiaoyan Ma$^{\star}$, Seohyun Lee$^{\star}$, Taejoon Kim$^{+}$, Christopher G. Brinton$^{\star}$ \\ $^{\star}$Purdue University, $^{+}$Arizona State University \\ $^{\star}$\{ma946, lee3296, cgb\}@purdue.edu, $^{+}$taejoonkim@asu.edu}





\begin{abstract}
Over-the-air federated learning (OTA-FL) improves communication efficiency by exploiting the superposition property of wireless channels, but this same property also creates a critical security vulnerability: the parameter server (PS) cannot access individual local updates, making it difficult to identify and exclude poisoned gradients. The challenge is further exacerbated under non-independent and identically distributed (Non-IID) training data, where benign gradient drift can closely resemble malicious updates.
In this paper, we propose a two-stage robust aggregation framework for defending against backdoor attacks in OTA-FL. Under our scheme, each client is first assigned a modality-aware multi-indicator trust score, where the specific indicators are selected according to the data modality (e.g., waveform, text, image) and model architecture to capture the most discriminative footprint of backdoor updates. Based on this score, the PS then performs trust-based multiple access (TBMA) to separate clients into trusted, suspicious, and malicious categories. Suspicious clients are further examined through PS-side layer-wise inspection and a longitudinal reputation mechanism.
Experimental results on several datasets demonstrate that the proposed methodology effectively suppresses stealthy backdoor attacks, including bounded-scaling attacks, Euclidean-constrained attacks, Cosine-constrained attacks, and Neurotoxin, while maintaining competitive main-task accuracy.
\end{abstract}

\keywords{Over-the-air Federated Learning, Backdoor Attack, Robust Aggregation Scheme, Trust Score, Layer-wise Inspection, Reputation Score}

\maketitle

\vspace{-10pt}

\section{Introduction}

Federated learning (FL) has become a popular distributed machine learning paradigm for collaborative model training across wireless devices. Over-the-Air FL (OTA-FL) further leverages the superposition property of wireless channels to achieve communication-efficient model aggregation,
offering a theoretically scale-free communication overhead for mobile edge networks. On the other hand, the open nature of wireless channels introduces critical security vulnerabilities: the parameter server (PS) cannot individually verify the correctness of local model updates, leaving OTA-FL systems highly susceptible to model poisoning attacks \cite{vulnerability,ma2025,wang2025mitigating}. 

A \textit{backdoor attack} is a form of model poisoning in which an adversary manipulates a subset of local clients to embed hidden malicious behavior into the globally aggregated model~\cite{Lee}. The attack is specifically designed to be stealthy, i.e., the compromised model maintains high accuracy on benign inputs while producing attacker-specified erroneous outputs whenever a predefined trigger pattern appears \cite{bagdasaryan2020backdoor}. More critically, adversaries can craft poisoned gradients according to specific constraints to further evade detection. For example, in a Euclidean-constrained attack, the malicious update is scaled so that its $\ell_2$ distance from the average benign gradient remains within a prescribed bound, while in a Cosine-constrained attack, the poisoned gradient is manipulated to maintain a high cosine similarity with the benign gradient direction. In both cases, the resulting malicious updates are geometrically indistinguishable from legitimate ones, rendering conventional norm- or direction-based detection methods ineffective. In OTA-FL, these challenges are further compounded by the fundamental nature of over-the-air aggregations: since the PS only receives the aggregated model update rather than the individual gradients of each client, any per-client inspection is infeasible \cite{no_indivudual}.

For traditional FL systems, extensive efforts have been devoted to defending against backdoor attacks, with two main categories of works. The first category consists of resource management strategies, including user grouping \cite{gruoping}, resource block allocation \cite{resource_block}, and hierarchical OTA-FL frameworks \cite{hierarchical}, which attempt to limit the influence of malicious participants by controlling wireless access. The second category comprises robust aggregation approaches that aim to exclude malicious gradients from the aggregation process or reduce the influence of malicious attacks, including well-known methods such as Krum \cite{Krum}, FoolsGold \cite{FoolsGold}, FLTrust \cite{FLTrust}, FLDetector \cite{FLDetector}, and FLAME \cite{Flame}. While these methods have shown effectiveness in standard FL settings, they require individual access to each client’s local model update, making them incompatible with OTA-FL. Moreover, they typically rely on a single metric to detect and filter malicious updates, which makes them inherently vulnerable to metric-constrained backdoor attacks, where a sophisticated adversary can tailor its poisoning strategy to evade the specific metric being monitored \cite{Shejwalkar2021Manipulating}. In this work, we consider the important open question of how to defend against backdoor attacks in OTA-FL.

\section{Related Works}
We review representative backdoor attack models and existing defense mechanisms that are most relevant to our work.

\subsection{Backdoor Attacks in FL and OTA-FL} \label{Backdoor_type}
In backdoor attacks, the adversary trains a poisoned local model and uploads it for global aggregation, with the goal of forcing attacker-chosen inputs to be mapped to a target label while maintaining normal behavior on benign inputs. Both conventional FL and OTA-FL are vulnerable to these attacks, which can be broadly categorized according to the adversary's capability:
\enlargethispage{\baselineskip}

\begin{itemize}[leftmargin=*, itemsep=1pt, topsep=1pt, parsep=0pt]
\item \textbf{Black-box attack:} In this type of attack, the adversary can only manipulate the local training dataset without accessing model parameters \cite{blackbackdoor}. Depending on whether poisoned samples are relabeled, black-box attacks can be broadly divided into dirty-label and clean-label attacks \cite{Lee}. Dirty-label attacks assign poisoned samples to the target class, as in distributed backdoor attack (DBA) \cite{xie2019dba}, which decomposes a global trigger into multiple local triggers across malicious clients. By contrast, clean-label attacks preserve the original labels and use carefully selected or crafted samples to implant the backdoor, as in edge-case-based attacks \cite{wang2020attack,zhang2023a3fl}, which exploit rare or carefully constructed samples to improve stealthiness and persistence.

    \item \textbf{White-box attack}: In this type of attack, the adversary can manipulate both the training data and local model parameters \cite{blackbackdoor}. After local training, the attacker modifies the model parameters to maximize impact \cite{bagdasaryan2020backdoor,li20233dfed,zhang2022neurotoxin} and/or reduce the deviation of backdoor updates \cite{wang2020attack,bagdasaryan2020backdoor,li20233dfed}. Model replacement attacks \cite{bagdasaryan2020backdoor} focus solely on attack impact by scaling the model updates and replacing the global model, and thus have been found relatively easy to filter out \cite{blackbackdoor}. To improve stealthiness, some works focus on \textit{constrained} backdoor attack models. For example, the constrain-and-scale attack \cite{bagdasaryan2020backdoor} constrains the training process itself to evade anomaly detection. The Neurotoxin attack \cite{zhang2022neurotoxin} projects the backdoor gradient onto the subspace unused by benign clients to achieve a durable attack.  
\end{itemize}
In this paper, we focus on the more challenging \textit{constrained white-box} backdoor attack setting, specifically for OTA-FL systems.

\subsection{Backdoor Defenses in OTA-FL}
Defenses against backdoor attacks have been extensively studied \cite{Krum,FoolsGold,FLTrust,FLDetector,Flame} for conventional FL setups. However, most existing defense strategies rely on the assumption that the PS can separately access each client’s local model update before aggregation, which as discussed does not hold in OTA-FL systems.
Indeed, due to the superposition property, the PS cannot directly inspect, compare, or selectively filter each client's model update prior to aggregation, which renders many defense mechanisms developed for conventional FL inapplicable to OTA-FL.

In this regard, a few studies have focused on designing robust aggregation mechanisms to enhance resilience against poisoning attacks. The best-effort voting (BEV) method for analog aggregation-based FL was proposed in \cite{BEV_SGD}. BEV instructs benign clients to transmit with maximum power to counteract the influence of malicious updates. Other approaches adopt client grouping and resource block allocation approaches to dilute the impact of corrupted gradients and improve robustness against adversarial participants \cite{group0,group1,group2}. The authors in \cite{dummy0,dummy1} propose to insert a specified number of dummy symbols into the updated local gradients to detect the presence of attackers. Finally, \cite{BC0,BC1} propose blockchain-based frameworks to mitigate poisoning attacks, employing ``reward-and-slash'' designs in which the clients propose candidate sets of benign and adversarial users and vote on them.

Although these methods enhance resilience in OTA-FL, \textit{they often rely on additional system resources, i.e., dedicated resource blocks and extra transmission symbols, or architectural modifications}. Ideally, a backdoor defense for OTA-FL will minimize added overhead and avoid disrupting the OTA aggregation process of benign clients. This motivates our design of \textit{detection-aware} schemes that introduce extra communication only for suspicious users.
The challenge of identifying adversarial devices is further amplified under non-independent and identically distributed (Non-IID) data distributions, where heterogeneous measurement distributions across local devices naturally induce gradient drift among benign clients. As a result, the PS must contend with the additional challenge of distinguishing between \textit{drifted benign updates} caused by statistical heterogeneity and \textit{carefully crafted malicious updates} that mimic similar deviation patterns.

\section{Summary of Contributions}
The aforementioned discussions reveal three key challenges in defending against backdoor attacks in OTA-FL systems. 
First, the PS cannot rely on access to individual client updates due to signal superposition in OTA-FL. Second, existing OTA-FL defenses often introduce additional communication or architectural complexity, thereby compromising system efficiency. Third, under Non-IID data distributions, benign gradient drift makes malicious updates harder to distinguish. To address these challenges, in this work, we propose \textbf{Trust-Then-Inspect (TTI)}, a two-stage robust aggregation framework for detecting backdoor attacks in OTA-FL systems. By confining additional communication overhead exclusively to suspicious clients, TTI preserves the communication efficiency of OTA aggregation for the majority of trusted participants while enabling fine-grained inspection only where it is warranted. 

\begin{figure}
    \centering
    \includegraphics[width=1\linewidth]{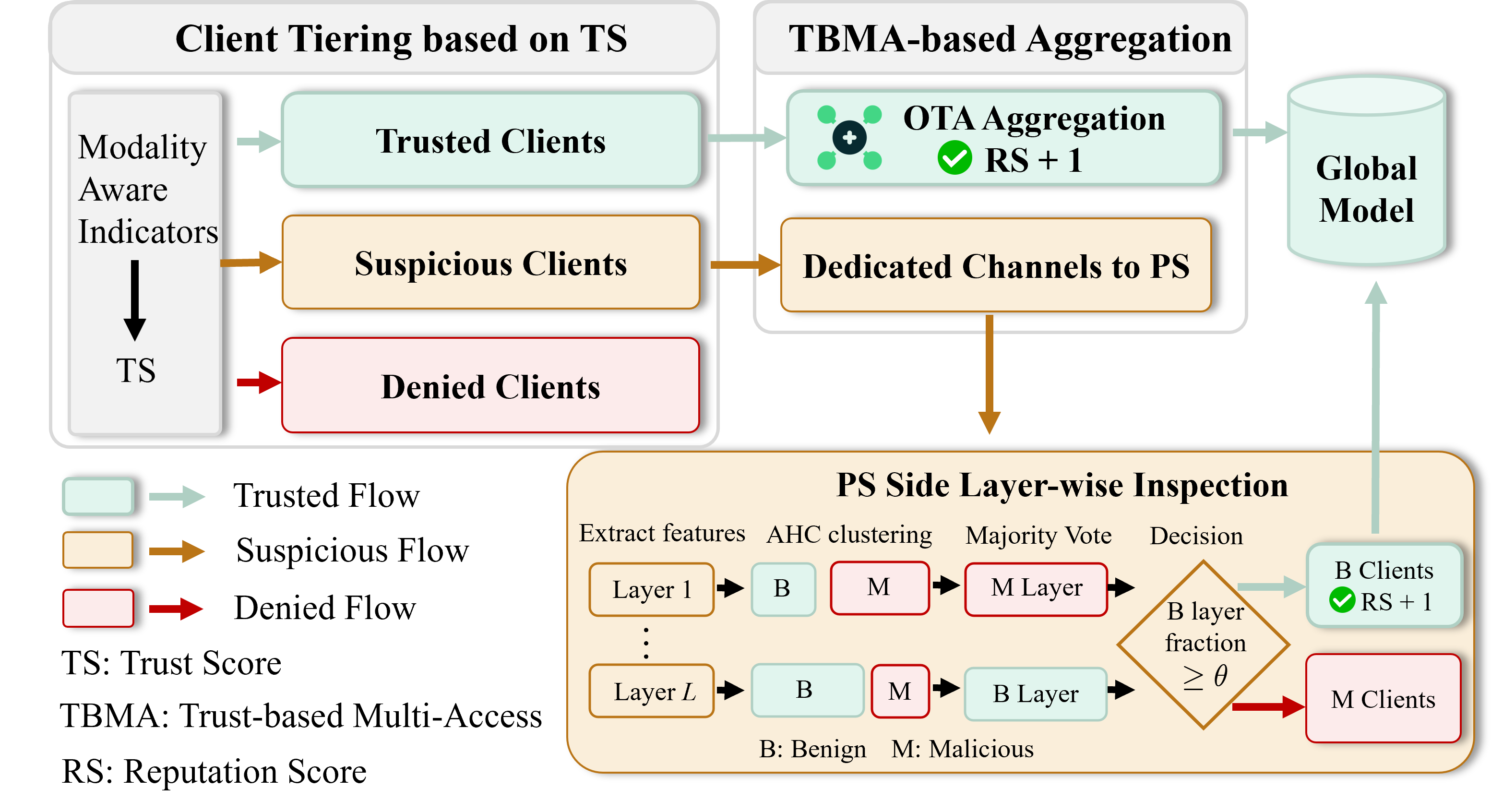}
    \captionsetup{font=small, width=\linewidth}
    \caption{Proposed two-stage robust aggregation framework  for detecting backdoor attacks in OTA-FL systems.}
    \label{Fig:systemflow}
\end{figure}

An overview of TTI is given in Fig. \ref{Fig:systemflow}. Referring to this figure, our specific contributions of this work are summarized as follows:
\begin{itemize}[leftmargin=*, itemsep=1pt, topsep=1pt, parsep=0pt]
    \item We develop a client-side, modality-aware, multi-indicator trust scoring mechanism for TTI. Specific indicators are selected within an overall trust-scoring framework based on task modality (e.g., waveform, image, text) to better capture the characteristic footprint of backdoor updates. Combiner weights across indicators are learned via Bayesian optimization. Unlike defenses relying on a single criterion, the proposed trust score provides more robustness against metric-constrained attacks in which adversaries craft poisoned gradients to circumvent any individual detection metric, as validated by our experiments.
    
    \item We design a PS-side layer-wise inspection module for TTI that performs anomaly detection at the granularity of individual neural network layers rather than at the global model level. This is motivated by our observation that metric-constrained malicious updates often appear benign under global metrics but exhibit anomalous behavior within specific layers. The proposed module extracts layer-wise gradient statistics and applies Agglomerative Hierarchical Clustering (AHC) followed by majority voting across layers to produce a robust final decision.
    
    \item We introduce a long-term reputation scoring (RS) mechanism that accumulates client behavioral evidence across training rounds. By maintaining a persistent per-client reputation score and applying a robust median absolute deviation (MAD)-based RS threshold, the PS progressively identifies malicious clients with increasing confidence. Moreover, as the RS values stabilize over time, the PS can increasingly rely on the RS-based filter to determine aggregation eligibility, thereby reducing the need for repeated trust scoring and layer-wise inspection and amortizing the long-term computational overhead of the defense framework.

    \item We conduct comprehensive experimental validations of the proposed TTI framework across multiple datasets, model architectures, attack models, and defense baselines. The results show that TTI consistently achieves substantially lower attack success rate (ASR) while maintaining competitive main task accuracy (MTA). Moreover, the experimental findings reveal that single-indicator tiering is vulnerable to its corresponding constrained attacks, whereas the proposed PS-side layer-wise inspection is crucial for detecting stealthy sparse backdoor updates. Comparisons with representative benchmarks further show that TTI attains robustness gains with only slight additional runtime overhead.
\end{itemize}

\section{Problem Setting}
\subsection{Network Architecture}
We consider an OTA-FL system where a PS orchestrates the training of a global model by coordinating a set of local clients. The communication between the PS and these clients is facilitated by a base station (BS). All clients use the same neural
network structure for training, which contains $S$ parameters (i.e., the number of model weights). Each local device $k$, $k=1,2,..., K$ has its own local dataset $\mathcal{D}_k=\lbrace (x_{k,i},y_{k,i}), \forall ~i \in \lbrace 1,2,...,D_k\rbrace \rbrace$, where $D_k$ is the number of training samples, $x_{k,i}$ is the input feature vector and $y_{k,i}$ is the corresponding ground-truth. In addition to benign clients participating in the OTA-FL process, there exists an adversary set $\mathcal{M}$ consisting of $M$ backdoor attackers. These attackers intentionally inject malicious model updates embedded with predefined trigger–target mappings, aiming to implant backdoor behaviors into the global model while preserving normal performance on clean data. Thereby in the system, we have $B=
K-M$ benign clients, and we use $\mathcal{B}$ to indicate the set of all benign clients in the system.
\subsection{Adversary Goals and Capabilities}
\textbf{(1) Adversarial Goals:} Each malicious
attacker in the adversary set $\mathcal{M}$ aims to manipulate the global model $\boldsymbol{w}$ to the poisoned model $\boldsymbol{w}_{M}$, which outputs an incorrect label $l_M$ for the adversary-chosen input set $\mathcal{I}_{M}$. We can formulate this as:
\begin{equation}
\boldsymbol{w}_M(x)=
\begin{cases}
l_M \neq \boldsymbol{w}(x), & \forall x \in \mathcal{I}_M, \\
\boldsymbol{w}(x), & \forall x \notin \mathcal{I}_M,
\end{cases}
\end{equation}
with $x$ representing the input of the model. In detail, the goals of backdoor attacks are two-fold: \textit{a) Effectiveness:} the poisoned model $\boldsymbol{w}_{M}$ predicts a specific incorrect label $l_M$  for any adversary-chosen input $ \forall x \in \mathcal{I}_M$, usually triggered by a pixel-trigger \cite{zhang2023a3fl} or a semantic trigger \cite{zhang2022neurotoxin}; \textit{b) Stealthiness:} the poisoned model $\boldsymbol{w}_{M}$ outputs the correct prediction on any non-adversary-chosen input $ \forall x \notin \mathcal{I}_M$ to maintain the main task accuracy. To achieve stealthiness, the poisoned model $\boldsymbol{w}_{M}$ should be indistinguishable from the benign models $\boldsymbol{w}$, in order to bypass the defense mechanism in the system.

\textbf{(2) Adversarial Capabilities:}  
 Detection-based defenses mainly rely on specific metrics to measure the distance between clients’ gradients and the global model, i.e., $dist_{metric}(\boldsymbol{w},\boldsymbol{w^{'}})$. Given that backdoor attackers have certain knowledge of the defense strategy, they can make their attacks highly stealthy by adapting to the specific defense. Previous attacks \cite{bagdasaryan2020backdoor,li20233dfed}
proposed to adapt the loss function to make the model more inconspicuous by adding a term $\mathcal{L}_{metric}$ that measures the deviation between the last global model and the attack model, e.g., Euclidean or Cosine distance \cite{bagdasaryan2020backdoor}, to guarantee that the malicious update remains close to the benign model in the parameter space, thereby reducing its detectability during the screening process. In the local training of malicious client $m \in \mathcal{M}$ , the loss of the attack model $\mathcal{L}_{m}$ can be designed based on
\begin{equation}
    \mathcal{L}_{m}=(1-\alpha)\mathcal{L}_{normal}+\alpha\mathcal{L}_{metric},
\end{equation}
where $\alpha \in [0,1]$ balances the weight of the effectiveness and the stealthiness of the attack and $\mathcal{L}_{normal}$ denotes the cross entropy loss function in normal
training on both clean and backdoor data. For example, in \cite{bagdasaryan2020backdoor}, the Cosine distance constrained backdoor attack is achieved by adding a Cosine constraint item into the loss function of backdoor training to constrain the cosine deviation between the backdoor and the global model, i.e.,
\begin{equation}
    \mathcal{L}_{m}=(1-\alpha)\mathcal{L}_{normal}+\alpha\mathcal{L}_{cosine}.
\end{equation}
The $\mathcal{L}_{cosine}$ represents
the cosine distance between the current backdoor model $\boldsymbol{w}_{M}^t$ and the last global model $\boldsymbol{w}^{t-1}$ distributed by the PS in this round. Here, $t \in [1,2,..,T]$ represents the index of communication round between local clients and the PS. Specifically, $\mathcal{L}_{cosine}$ can be computed as 
\begin{equation}
    \mathcal{L}_{cosine} = 1 - \frac{\langle \boldsymbol{w}^{t-1}, \boldsymbol{w}_{M}^t \rangle}{|\boldsymbol{w}^{t-1}| |\boldsymbol{w}_{M}^t|}.
\end{equation}
Here, $\langle \boldsymbol{w}^{t-1}, \boldsymbol{w}_M^t \rangle$ denotes 
the inner product between the last global model and the attack model, 
and $|\boldsymbol{w}^{t-1}|$, $|\boldsymbol{w}_M^t|$ denote their  respective $\ell_2$ norms. A smaller $\mathcal{L}_{cosine}$ indicates a higher cosine similarity between $\boldsymbol{w}_M^t$ and 
$\boldsymbol{w}^{t-1}$, meaning that the attack model is more directionally aligned with the global model and thus harder to detect.

\section{Two-Stage Robust Aggregation Framework} \label{sec:framework}
In this paper, we aim to design a detection-based defense for OTA-FL that can thwart well-crafted backdoor attacks, where sophisticated adversaries carefully manipulate malicious gradients to evade detection. At a high level, the proposed defense should satisfy three key properties: \textbf{a) Effectiveness:} it should reliably identify and filter malicious gradients, even when attackers constrain their updates to resemble benign statistics; \textbf{b) Utility:} it should preserve the convergence and accuracy of the main task while introducing minimal performance degradation; and \textbf{c) Generalizability:} it should remain applicable across diverse datasets, model architectures, federated learning settings, and attack strategies. Based on these considerations, we propose the following two-stage robust aggregation scheme for OTA-FL systems.

\subsection{Stage I: Client Tiering via Trust Score}
The key idea behind trust score designing is to perform a lightweight preliminary screening at the client side before resorting to more sophisticated detection at the PS. In OTA-FL systems, clients are typically resource-constrained devices with limited computational capabilities. Therefore, rather than imposing a complex detection mechanism uniformly on all clients, we first employ a simple trust score computation that relies on basic statistical indicators to quickly categorize clients into different trust tiers. This coarse-grained tiering serves as an efficient pre-filter: clients deemed clearly trustworthy or clearly malicious are immediately accepted or rejected, respectively, while only the ambiguous suspicious clients are forwarded to the PS for further layer-wise inspection in Stage~II. By concentrating the expensive fine-grained analysis on a small subset of clients, this two-stage design significantly reduces the overall computational burden while maintaining strong detection capability.
\subsubsection{Modality-Aware Indicators for Trust Score Calculation.}~\label{indicatordesign}
The Stage I trust score is computed from multiple indicators, while the exact indicator set may vary across data modalities and model architectures. This is because backdoor attacks can leave different structural footprints in model updates for different tasks. In this stage, each client computes a set of lightweight local indicators and reports them to the PS, which then evaluates the client’s trustworthiness through a composite trust score. Using multiple indicators makes the score more difficult for a well-crafted adversary to evade than any single detection metric. We first introduce three representative geometric indicators, which can be further complemented by task-specific indicators when needed.

\textbf{(1) Temporal Direction Alignment (TDA)} measures the directional consistency between client $k$'s local update and the global model via cosine similarity:
\begin{equation}\label{eq:tda}
    \omega_k^t = \frac{\langle \Delta_k^t,\, \boldsymbol{w}^{t-1} \rangle}{\|\Delta_k^t\|_2 \, \|\boldsymbol{w}^{t-1}\|_2},
\end{equation}
where $\Delta_k^t = \boldsymbol{w}_{k}^t - \boldsymbol{w}^{t-1}$ is the local model update of client $k$, $\boldsymbol{w}_{k}^t$ is the locally updated model for client $k$ at round $t$ and $\boldsymbol{w}^{t-1}$ is the global model broadcast at the beginning of round $t$.

\textbf{(2) Relative $\ell_2$ Norm} captures abnormal magnitude deviations of client $k$'s update relative to the global model:
\begin{equation}\label{eq:rel_l2}
    \mathrm{rel\ell_2}_k(t) = \frac{\|\Delta_k^t\|_2}{\|\boldsymbol{w}^{t-1}\|_2}.
\end{equation}
A disproportionately large relative norm suggests that the client's update deviates significantly from the expected magnitude.

\textbf{(3) Spikiness} quantifies the concentration of gradient energy in the top coordinates, detecting whether a small subset of parameters carries an anomalously large fraction of the total update energy:
\begin{equation}\label{eq:spikiness}
    \xi_k(t) = \frac{\|\Delta_k^{t,\,\mathrm{top\text{-}1\%}}\|_2^2}{\|\Delta_k^t\|_2^2},
\end{equation}
where $\Delta_k^{t,\,\mathrm{top\text{-}1\%}}$ denotes the entries of $\Delta_k^t$ corresponding to the top $1\%$ coordinates by absolute value. A high spikiness ratio is a hallmark of backdoor attacks that concentrate perturbations in a few critical parameters.
\enlargethispage{\baselineskip}

Let $\phi_k^{(i)}(t)$ denote the $i$-th indicator for client $k$ at round $t$, where $i \in \{1,2,\dots,I_m\}$ and $I_m$ denotes the number of indicators. Since these raw indicators may have different scales and ranges, we apply per-indicator monotone transformations to map them into $[0,1]$. 
For the common geometric indicators introduced above, the TDA similarity defined in \eqref{eq:tda} is linearly rescaled as
\begin{equation}\label{eq:tda_transform}
    \bar{\phi}_k^{(1)}(t) = \frac{\omega_k(t)+1}{2}.
\end{equation}
The relative $\ell_2$ norm defined in \eqref{eq:rel_l2} is transformed via a shifted hyperbolic tangent function:
\begin{equation}\label{eq:rel_l2_transform}
    \bar{\phi}_k^{(2)}(t) = 1 - \tanh\!\Big(\alpha \cdot \max\!\big(0,\; \mathrm{rel\ell_2}_k(t) - r_0\big)\Big),
\end{equation}
where $r_0$ is a baseline threshold and $\alpha$ controls the steepness of the penalty. The spikiness defined in \eqref{eq:spikiness} is transformed via a polynomial penalty:
\begin{equation}\label{eq:spiky_transform}
    \bar{\phi}_k^{(3)}(t) = 1 - \Big[\max\!\big(0,\; \xi_k(t) - s_0\big)\Big]^\gamma,
\end{equation}
where $s_0$ is a tolerance threshold and $\gamma$ controls the decay rate.

For the remaining modality-specific indicators, suitable monotone transformations are applied so that all normalized indicators satisfy $\bar{\phi}_k^{(i)}(t)\in[0,1]$, where a larger value indicates more benign behavior. Accordingly, the composite trust score is computed in a unified form as
{ \setlength{\abovedisplayskip}{1pt}
\setlength{\belowdisplayskip}{2pt}
\begin{equation}\label{eq:trust_score}
    TS_k^t = \sum_{i=1}^{I_m} \beta_i \, \bar{\phi}_k^{(i)}(t),
\end{equation}}
where the weight vector $\boldsymbol{\beta} = [\beta_1,\beta_2,\dots,\beta_{I_m}]^T$ satisfies $\sum_{i=1}^{I_m} \beta_i = 1$ and $\beta_i \geq 0$, ensuring $TS_k^t\in[0,1]$.

\subsubsection{Weight Optimization for Each Indicator.}~A critical design challenge in \eqref{eq:trust_score} is the selection of the weight vector $\boldsymbol{\beta}$. Manually tuning these weights is impractical, as the optimal combination depends on the specific attack strategy, data distribution, and model architecture. To address this, we formulate the weight design as a black-box optimization problem and employ Bayesian Optimization (BO) to automatically learn $\boldsymbol{\beta}$. Specifically, we define an evaluation score that jointly captures the defense's ability to maintain model accuracy while suppressing backdoor attacks:
\begin{equation}\label{eq:bo_objective}
    \mathcal{J}(\boldsymbol{\beta}) = \mathrm{Accuracy}(\boldsymbol{\beta}) - \lambda \cdot \mathrm{ASR}(\boldsymbol{\beta}),
\end{equation}
where $\mathrm{Accuracy}(\boldsymbol{\beta})$ is the global model's test accuracy and $\mathrm{ASR}(\boldsymbol{\beta})$ represents the attack success rate (ASR), both obtained by running a full OTA-FL simulation with the trust score weights set to $\boldsymbol{\beta}$. The trade-off parameter $\lambda > 0$ controls the relative penalty on the ASR. The goal is to find:
\begin{equation}\label{eq:bo_goal}
    \boldsymbol{\beta}^{\star} = \arg\max_{\boldsymbol{\beta} \in \mathcal{B}} \; \mathcal{J}(\boldsymbol{\beta}), \quad \mathcal{B} = \left\{ \boldsymbol{\beta} \in \mathbb{R}^{I_m}_{\geq 0} \;\middle|\; \sum_{i=1}^{I_m} \beta_i = 1 \right\}.
\end{equation}

Since each evaluation of $\mathcal{J}(\boldsymbol{\beta})$ requires running a complete multi-round OTA-FL simulation, the objective function is expensive to evaluate and lacks a closed-form expression. BO is particularly well-suited for this setting, as it constructs a probabilistic surrogate model to guide the search with minimal function evaluations. The BO procedure summarized in Algorithm~\ref{alg:bo_weights}, is used only as a one-time offline calibration step before deployment. Specifically, it is carried out in a development-stage validation environment under representative attack conditions, and the resulting weight vector $\boldsymbol{\beta}^{\star}$ is then fixed for subsequent online deployment. The BO search begins by sampling candidate weight vectors from a symmetric Dirichlet distribution over the $I_m$-dimensional probability simplex, i.e., $\mathrm{Dirichlet}(\mathbf{1}_{I_m})$, and evaluating each candidate through a full OTA-FL simulation. Using the collected $(\boldsymbol{\beta}, \mathcal{J})$ pairs, a Gaussian Process (GP) surrogate model is fitted to capture both the expected objective value and the associated uncertainty over the search space. The next candidate is then selected by maximizing the Expected Improvement (EI) acquisition function, which balances exploitation of promising regions and exploration of uncertain regions. After $N_{\mathrm{iter}}$ iterations, the weight vector with the highest observed evaluation score is returned as $\boldsymbol{\beta}^{\star}$. Once obtained, $\boldsymbol{\beta}^{\star}$ is fixed and used throughout online OTA-FL training, introducing no additional computational overhead at runtime.

\begin{algorithm}[t]
\caption{Bayesian Optimization for Trust Score Weights}
\label{alg:bo_weights}
\begin{algorithmic}[1]
\Statex \textbf{Input:} Number of initial samples $N_{\mathrm{init}}$, total BO iterations $N_{\mathrm{iter}}$, trade-off parameter $\lambda$, number of indicators $I_m$
\Statex \textbf{Output:} Optimal weight vector $\boldsymbol{\beta}^{\star}$
\Statex \textbf{--------- Phase 1: Initial Sampling ---------}
\For{$n = 1, \dots, N_{\mathrm{init}}$}
    \State Sample $\boldsymbol{\beta}^{(n)} \sim \mathrm{Dirichlet}(\mathbf{1}_{I_m})$
    \State Run OTA-FL simulation with weights $\boldsymbol{\beta}^{(n)}$
    \State Record corresponding evaluation
score $\mathcal{J}^{(n)}$ based on \eqref{eq:bo_objective}
\EndFor
\State Initialize dataset $\mathcal{D}_{\beta} = \{(\boldsymbol{\beta}^{(n)}, \mathcal{J}^{(n)})\}_{n=1}^{N_{\mathrm{init}}}$
\Statex \textbf{--------- Phase 2: Iterative Refinement ---------}
\For{$n = N_{\mathrm{init}} + 1, \dots, N_{\mathrm{init}} + N_{\mathrm{iter}}$}
    \State Fit a Gaussian Process (GP) surrogate model on $\mathcal{D}_{\beta}$ 
    \State Select next candidate $\boldsymbol{\beta}^{(n)} = \arg\max_{\boldsymbol{\beta} \in \mathcal{B}} \mathrm{EI}(\boldsymbol{\beta})$ 
    \State Run OTA-FL simulation with trust weights $\boldsymbol{\beta}^{(n)}$
    \State Record corresponding evaluation score $\mathcal{J}^{(n)}$ based on \eqref{eq:bo_objective}
    \State Update dataset $\mathcal{D}_{\beta} \leftarrow \mathcal{D}_{\beta} \cup \{(\boldsymbol{\beta}^{(n)}, \mathcal{J}^{(n)})\}$
\EndFor
\Statex \textbf{--------- Phase 3: Output ---------}
\State $\boldsymbol{\beta}^{\star} = \arg\max_{(\boldsymbol{\beta}, \mathcal{J}) \in \mathcal{D}_{\beta}} \mathcal{J}$
\State \Return $\boldsymbol{\beta}^{\star}$
\end{algorithmic}
\end{algorithm}
\subsubsection{Client Tiering Strategy.}~ \label{Tier_Strategy}
Based on the trust score, clients are partitioned into three tiers. Two strategies are supported: \textit{threshold-based} and \textit{proportion-based} tiering. In threshold-based tiering, two thresholds $\tau_{\mathrm{high}}$ and $\tau_{\mathrm{low}}$ ($\tau_{\mathrm{high}} > \tau_{\mathrm{low}}$) are used:
{ \setlength{\abovedisplayskip}{2pt}
\setlength{\belowdisplayskip}{2pt}
\begin{equation}\label{eq:tiering_threshold}
    \mathrm{Tier}_k^t =
    \begin{cases}
        \textit{Trusted},    & \text{if }~ TS_k^t \geq \tau_{\mathrm{high}}, \\
        \textit{Suspicious}, & \text{if }~ \tau_{\mathrm{low}} \leq TS_k^t < \tau_{\mathrm{high}}, \\
        \textit{Malicious},  & \text{if } ~TS_k^t < \tau_{\mathrm{low}}.
    \end{cases}
\end{equation}}
Alternatively, in proportion-based tiering, clients are ranked by their trust scores in descending order and partitioned according to predefined fractions $(p_T, p_S, p_M)$ with $p_T + p_S + p_M = 1$. Specifically, the top $p_T$ fraction of clients are classified as \textit{Trusted}, the next $p_S$ fraction as \textit{Suspicious}, and the bottom $p_M$ fraction as \textit{Malicious}.  
Furthermore, clients classified as trusted proceed directly to aggregation through OTA process. Clients in the suspicious tier are forwarded to Stage~II for further layer-wise inspection through individual uploading channel and clients classified as malicious are immediately excluded from the aggregation.

\subsection{Stage II: PS Side Layer-wise Inspection}

The suspicious clients identified in Stage~I occupy an ambiguous zone: their trust scores are neither high enough to be directly trusted nor low enough to be outright rejected. These clients may include either \textit{drifted benign clients}, whose updates deviate from the population due to non-IID data heterogeneity or noisy local training, or \textit{well-crafted malicious clients} whose metric-constrained backdoor updates are designed to appear plausible at the global level. The goal of Stage~II is to distinguish between these two cases through fine-grained, layer-wise inspection at the PS.
\textit{The key insight is that while malicious model updates may remain inconspicuous at the global level, they commonly exhibit anomalous behavior within particular layers of the network.} For instance, backdoor attacks often concentrate their perturbations in the final classifier layers while leaving earlier feature extraction layers largely unaffected. By inspecting each layer independently, the PS can detect such localized anomalies that would otherwise be masked by whole-model statistics \cite{LWG}.

\subsubsection{Layer-wise Feature Extraction and Clustering.}~
For each suspicious client $j \in \mathcal{S}_t$, where $\mathcal{S}_t$ represents the suspicious user set in communication round $t$, the PS decomposes its model update $\Delta_j^t$ into $L$ layer-wise gradient vectors $\{\mathbf{g}_{j,\ell}^t\}_{\ell=1}^{L}$. The trusted aggregate $\bar{\mathbf{g}}_{\mathrm{ref},\ell}^t = \frac{1}{|\mathcal{T}_t|} \sum_{k \in \mathcal{T}_t} \mathbf{g}_{k,\ell}^t$ serves as the benign reference for each layer, where $\mathcal{T}_t$ represent the set of trusted clients in communication round $t$. For each client in suspicious set $\mathcal{S}_t$, PS extracts a multi-dimensional feature vector characterizing the gradient structure at layer $\ell$. Specifically, for each candidate $j$, the following features are computed from the layer gradient $\mathbf{g}_{j,\ell}^t$:
\begin{itemize}[leftmargin=*, itemsep=1pt, topsep=1pt, parsep=0pt]
    \item \textbf{Sign structure:} the positive count, negative count, and zero count of gradient entries, capturing the sign distribution of the update;
    \item \textbf{Distribution shape:} the kurtosis and skewness of the gradient entries, measuring the sparsity (tailedness) and asymmetry of the distribution, respectively;
    \item \textbf{Inter-client distance:} the mean pairwise Euclidean distance ($D_{\mathrm{mean}}$) to all other candidates' gradients at this layer;
    \item \textbf{Reference deviation:} the $\ell_1$ deviation from the trusted reference gradient $\bar{\mathbf{g}}_{\mathrm{ref},\ell}^t$, detecting small but widespread modifications;
    \item \textbf{Magnitude:} the $\ell_2$ norm of the gradient, capturing localized large changes;
    \item \textbf{Directional similarity:} the angle between the gradient and the reference, computed as $\!\big(\langle \mathbf{g}_{j,\ell}^t, \bar{\mathbf{g}}_{\mathrm{ref},\ell}^t \rangle / (\|\mathbf{g}_{j,\ell}^t\|_2 \|\bar{\mathbf{g}}_{\mathrm{ref},\ell}^t\|_2)\big)$.
\end{itemize}

Using these features, the PS performs \textit{Agglomerative Hierarchical Clustering (AHC)} to partition the candidates into two clusters.

\subsubsection{Benign Cluster Identification.}~
After AHC produces two clusters $\mathcal{C}_1^{(\ell)}$ and $\mathcal{C}_2^{(\ell)}$ at layer $\ell$, the PS must determine which cluster is more likely to be benign. This decision is made based on four cluster-level statistics:
\begin{itemize}[leftmargin=*, itemsep=1pt, topsep=1pt, parsep=0pt]
    \item $D_{\mathrm{mean}}$: mean pairwise distance within the cluster (internal cohesion);
    \enlargethispage{\baselineskip}
    \item $\mathrm{SD}_{\mathrm{mean}}$: mean of per-coordinate standard deviations within the cluster (internal variability);
    \item $\mathrm{Dev}_{\mathrm{mean}}$: mean absolute deviation from the trusted reference (alignment with benign behavior);
    \item $\mathrm{RS}_{\mathrm{sum}}$: sum of the historical reputation scores of the clients whose layer $\ell$ updates are assigned to the cluster.
\end{itemize}


The benign cluster is selected as the one that, relative to the other cluster, satisfies either of the following conditions: (i) it has lower deviation from the trusted reference $(\mathrm{Dev}_{\mathrm{mean}})$ and comparable or higher accumulated reputation $(\mathrm{RS}_{\mathrm{sum}})$, even if its internal distance $(\mathrm{D}_{\mathrm{mean}})$ and internal variability $(\mathrm{SD}_{\mathrm{mean}})$ are larger; or (ii) it has both lower internal distance $(\mathrm{D}_{\mathrm{mean}})$ and lower deviation from the trusted reference $(\mathrm{Dev}_{\mathrm{mean}})$, together with comparable or higher accumulated reputation $(\mathrm{RS}_{\mathrm{sum}})$. In one word, the benign cluster is selected by prioritizing alignment with the trusted reference and historical reliability over internal consistency, since benign updates may exhibit high variability under non-IID data, while malicious updates tend to be more tightly clustered. Then clients assigned to the selected benign cluster at layer $\ell$ are marked as benign layer.

\subsubsection{Majority Vote and Final Decision.}~After processing all $L$ layers, each suspicious client $j$ receives a binary pass/fail verdict at each layer. The PS then computes the benign-layer fraction of client $j$ as
\begin{equation} \label{eq:benignfraction}
f_j=\frac{1}{L}\sum_{\ell=1}^{L}\mathbf{1}\{j \text{ passes layer } \ell\},
\end{equation}
where $\mathbf{1}\{\cdot\}$ denotes the indicator function. Then PS applies a majority vote:
\begin{equation}\label{eq:layer_vote}
    \mathrm{Decision}_j^t =
    \begin{cases}
        \textit{Accept},  & f_j \geq \rho, \\
        \textit{Reject},  &  f_j < \rho,
    \end{cases}
\end{equation}
where $\rho \in (0,1)$ is the benign layer fraction threshold. Accepted suspicious clients are promoted to the trusted tier and participate in the final aggregation, and rejected clients are excluded.
\FloatBarrier
 \begin{algorithm}[!t]
\caption{Trust-Then-Inspect (TTI) Framework for OTA-FL}
\label{alg:tti}
\begin{algorithmic}[1]
\Statex \textbf{Input:} Number of clients $K$, communication rounds $T$, tiering fractions $(p_T, p_S, p_M)$, benign layer fraction threshold $\rho$, warm-up rounds $T_{\mathrm{warm}}$
\Statex \textbf{Output:} Global model $\boldsymbol{w}^T$
\State Initialize global model $\boldsymbol{w}^0$ and set $\mathrm{RS}_k = 0$ for all $k$
\For{$t = 1, \dots, T$}
    \State PS broadcasts $\boldsymbol{w}^{t-1}$ to all $K$ clients
    \For{each client $k = 1, \dots, K$ \textbf{in parallel}}
        \State Perform local SGD on $\mathcal{D}_k$ to obtain $\boldsymbol{w}_k^t$
        \State Compute local indicators and send them to PS
    \EndFor
     \Statex \hspace{\algorithmicindent}
     \textbf{--------- Step 1: Proportion-based Client Tiering ---------}
    \State PS calculates $TS_k^t$ based on \eqref{eq:trust_score} and ranks clients in descending order; Top $p_T$ fraction $\to \mathcal{T}_t$, next $p_S$ $\to \mathcal{S}_t$, bottom $p_M$ $\to \mathcal{M}_t$
    \Statex \hspace{\algorithmicindent} \textbf{--------- Step 2: TBMA-based OTA Aggregation ---------}
    \For{each client $k \in \mathcal{T}_t$}
        \State Transmit local gradients through OTA
    \EndFor
    \State PS obtains trusted aggregate $\bar{\mathbf{g}}_{\mathrm{ref}}^t$ as benign reference
    \For{each client $j \in \mathcal{S}_t$}
        \State Upload $\Delta_j^t$ via individual transmit link
    \EndFor
    \Statex \hspace{\algorithmicindent} \textbf{--------- Step 3: Layer-wise Inspection ---------}
    \For{each suspicious client $j \in \mathcal{S}_t$}
        \State Decompose $\Delta_j^t$ into layer-wise gradients $\{\mathbf{g}_{j,\ell}^t\}_{\ell=1}^{L}$
        \For{each layer $\ell = 1, \dots, L$}
            \State Extract features; perform AHC into 2 clusters
            \State Identify benign cluster 
        \EndFor
        \State Compute benign layer fraction based on \eqref{eq:benignfraction}
        \If{$f_j \geq \rho$}
            \State Accept: $\mathcal{T}_t \leftarrow \mathcal{T}_t \cup \{j\}$
        \Else
            \State Reject: $\mathcal{M}_t \leftarrow \mathcal{M}_t \cup \{j\}$
        \EndIf
    \EndFor
    \Statex \hspace{\algorithmicindent} \textbf{--------- Step 4: RS-based Filtering (after $T_{\mathrm{warm}}$) ---------}
    \If{$t > T_{\mathrm{warm}}$}
        \State Compute $\tilde{\mathrm{RS}}$, $\mathrm{MAD}_{\mathrm{RS}}$ over all clients
        \State Remove clients with $\mathrm{RS}_k < \tilde{\mathrm{RS}} - \mathrm{MAD}_{\mathrm{RS}}$ from $\mathcal{T}_t$
    \EndIf
    \State Update $\mathrm{RS}_k \leftarrow \mathrm{RS}_k + 1$ for all $k$ in final participant set
    \Statex \hspace{\algorithmicindent} \textbf{--------- Step 5: Global Model Update ---------}
    \State Compute the final global model from final participant set
\EndFor
\State \Return $\boldsymbol{w}^T$
\end{algorithmic}
\end{algorithm}
\subsubsection{Reputation Score (RS).}~
To incorporate historical behavior into the inspection process, the PS maintains a persistent reputation score $\mathrm{RS}_k$ for each client $k$. At the end of each round, every client that participates in the final aggregation receives an increment 
$\mathrm{RS}_k \leftarrow \mathrm{RS}_k + 1$. Clients that are consistently accepted accumulate higher RS values over time, while malicious clients that are frequently rejected maintain low RS.
After a warm-up period of $T_{\mathrm{warm}}$ rounds, the PS applies an additional RS-based filter as a final safeguard. Let $\tilde{\mathrm{RS}}$ and $\mathrm{MAD}_{\mathrm{RS}}$ denote the median and MAD of all clients' reputation scores, respectively. A client is retained in the final participant set only if:
\begin{equation}\label{eq:rs_filter}
    \mathrm{RS}_k \geq \tilde{\mathrm{RS}} - \mathrm{MAD}_{\mathrm{RS}}.
\end{equation}
This RS-based filter provides a long-term defense mechanism: even if a sophisticated attacker occasionally evades the layer-wise inspection in a single round, its persistently low RS will eventually lead to its exclusion.  Moreover, as the RS values stabilize over sufficient training rounds, the PS can progressively rely on the RS filter alone to determine client eligibility for aggregation, bypassing the computationally expensive trust scoring and layer-wise inspection stages. This significantly reduces the long-term computational overhead of the defense framework, effectively amortizing the cost of the initial inspection rounds over the entire training process.

\begin{table*}[!t]
\centering
\caption{Dataset-specific Settings.}
\label{tab:indicator_config}
\setlength{\tabcolsep}{4pt}
\renewcommand{\arraystretch}{1.05}
\begin{tabular}{>{\centering\arraybackslash}m{1.8cm}|
                >{\centering\arraybackslash}m{1.3cm}|
                >{\centering\arraybackslash}m{1.2cm}|
                >{\centering\arraybackslash}m{2cm}|
                >{\centering\arraybackslash}
                m{6.5cm}|
                >{\centering\arraybackslash}
                 m{0.8cm}|
                 >{\centering\arraybackslash}
                 m{0.5cm}|
                 >{\centering\arraybackslash}
                 m{1.3cm}
                 }
                
\hline
\textbf{Dataset} & \textbf{Modality} & \textbf{Task} & \textbf{Architecture} & \textbf{Stage-I Trust Indicators} & $T_{\mathrm{warm}}$ & $T$ &$\rho$ in \eqref{eq:layer_vote} \\
\hline
RML2016.10A & Waveform & 11-class & 1D CNN &
Rel$\ell_2$ norm, spikiness, and consistency indicators& 10 &100 &0.6\\
\hline
AG News & Text & 4-class & Text CNN &
Rel$\ell_2$ norm, spikiness, and text-specific indicators & 10 & 100&0.6\\
\hline
CIFAR-10 & Image & 10-class & CNN / ResNet9 &
TDA, rel$\ell_2$ norm, and spikiness &20& 100&0.6\\
\hline
\end{tabular}
\vspace{2mm}
\end{table*}

\subsection{Overall Process for Proposed Framework}
Before introducing the algorithm, we first clarify the attack model considered in this paper. We assume that a compromised client has full control over its local training process, including its local dataset and model parameters, and can therefore generate carefully crafted malicious updates. However, the attacker is not assumed to arbitrarily spoof the Stage I feedback independently of the actual local update. Instead, the reported indicators are assumed to be derived from the client’s true update and delivered through a trusted or authenticated reporting interface. Under this assumption, an adversary may still attempt to craft its update so that these indicators appear benign, but it cannot freely report an inconsistent high-trust statistic vector unrelated to the actual update. Therefore, Stage I is interpreted as a coarse-grained screening mechanism rather than a standalone security guarantee, while Stage II layer-wise inspection and RS-based filtering provide additional safeguards against evasive malicious clients. The overall TTI procedure is summarized in Algorithm~\ref{alg:tti}, and its main steps are briefly outlined as follows.

\textbf{Step 1 (Trust Scoring \& Tiering):} Each client computes its local indicators and feeds them back to the PS. Based on the resulting trust scores, the PS partitions all clients into the trusted set $\mathcal{T}_t$, suspicious set $\mathcal{S}_t$, and malicious set $\mathcal{M}_t$.
\enlargethispage{\baselineskip}

\textbf{Step 2 (TBMA-based Aggregation):} Clients in $\mathcal{T}_t$ participate in OTA aggregation to form a trusted reference, while clients in $\mathcal{S}_t$ upload their updates individually for further inspection. Clients in $\mathcal{M}_t$ are directly excluded.

\textbf{Step 3 (Layer-wise Inspection):} The PS performs layer-wise inspection on each suspicious client in $\mathcal{S}_t$ and determines its accept/reject decision according to the benign layer fraction.

\textbf{Step 4 (RS-based Filtering):} After the warm-up period $T_{\mathrm{warm}}$, an additional RS-based filter is applied to the accepted clients, where only those satisfying $\mathrm{RS}_k \geq \tilde{\mathrm{RS}}-\mathrm{MAD}_{\mathrm{RS}}$ are retained. The RS values of the final participants are then updated.

\textbf{Step 5 (Global Model Update):} The global model is updated using the updates from the final participant set.

 \begin{table*}[!t]
\centering
\setlength{\tabcolsep}{3pt}
\renewcommand{\arraystretch}{1.15}
\caption{Main Task Accuracy (MTA) and Attack Success Rate (ASR) of the proposed TTI framework under Non-IID settings. We see that our method substantially improves the ASR while preserving MTA compared to no defense.}
\label{tab:exp1_effectiveness}
\renewcommand{\arraystretch}{1.2}
\begin{tabular}{l|cc|cc|cc|cc}
\hline
\multirow{2}{*}{\textbf{Defense and Attack Strategy}} & \multicolumn{2}{c|}{\textbf{RML2016.10A}} & \multicolumn{2}{c|}{\textbf{AGNews}} & \multicolumn{2}{c|}{\textbf{CIFAR-10 (CNN)}} & \multicolumn{2}{c}{\textbf{CIFAR-10 (ResNet 9)}}  \\

\cline{2-9}
 & MTA ($\uparrow$) & ASR ($\downarrow$) & 
 MTA ($\uparrow$) & ASR ($\downarrow$) &
 MTA ($\uparrow$) & ASR ($\downarrow$)&
MTA ($\uparrow$) & ASR ($\downarrow$)\\
\hline
\hline
No Attack (All Clients Participate) & 80.07 $ {\scriptstyle \pm \ 0.34}$ & -- & 90.38 ${\scriptstyle \pm \ 0.23}$ & -- & 76.87 ${\scriptstyle \pm \ 1.09}$ & -- & 82.78 ${\scriptstyle \pm \ 0.98}$ & -- \\
\hline
\rowcolor{gray!10}
No Defense + Bounded-scaling & 79.94 ${\scriptstyle \pm \ 2.30}$  & 99.97 ${\scriptstyle \pm \ 0.02}$ & 90.82 ${\scriptstyle \pm \ 1.56}$ & 100.00 ${\scriptstyle \pm \ 0.00}$ & 75.42 ${\scriptstyle \pm \ 2.03}$ & 95.54 ${\scriptstyle \pm \ 0.69}$ & 82.31 ${\scriptstyle \pm \ 2.31}$ & 98.14 ${\scriptstyle \pm \ 0.84}$\\
\rowcolor{cyan!10}
No Defense + Euc-constrained & 79.03 ${\scriptstyle \pm \ 3.42}$& 84.11 ${\scriptstyle \pm \ 1.13}$ & 87.51 ${\scriptstyle \pm \ 3.90}$ & 100.00 ${\scriptstyle \pm \ 0.00}$ & 79.93 ${\scriptstyle \pm \ 3.68}$ & 88.54 ${\scriptstyle \pm \ 0.79}$& 78.03 ${\scriptstyle \pm \ 1.27}$& 81.77 ${\scriptstyle \pm \ 3.11}$\\
\rowcolor{blue!8}
No Defense + Cos-constrained & 77.65 ${\scriptstyle \pm \ 1.84}$ & 69.56 ${\scriptstyle \pm \ 3.59}$& 90.51 ${\scriptstyle \pm \ 0.06}$ & 100.00  ${\scriptstyle \pm \ 0.00}$&  75.64 ${\scriptstyle \pm \ 0.43}$&  83.14 ${\scriptstyle \pm \ 0.67}$ & 77.84 ${\scriptstyle \pm \ 0.21}$ & 70.01 ${\scriptstyle \pm \ 2.47}$ \\
\rowcolor{red!8}
No Defense + Neurotoxin & 79.90 ${\scriptstyle \pm \ 0.19}$ & 99.89 ${\scriptstyle \pm \ 0.06}$ & 90.68 ${\scriptstyle \pm \ 0.41}$ & 68.45 ${\scriptstyle \pm \ 1.98}$ & 77.51 ${\scriptstyle \pm \ 1.56}$ & 76.56 ${\scriptstyle \pm \ 2.30}$ & 82.58 ${\scriptstyle \pm \ 1.09}$& 77.85 ${\scriptstyle \pm \ 2.31}$\\
\hline
\rowcolor{gray!10}
Proposed TTI + Bounded-scaling& 78.16 ${\scriptstyle \pm \ 0.58}$& 12.65 ${\scriptstyle \pm \ 1.26}$ & 86.88 ${\scriptstyle \pm \ 0.67}$& 22.05 ${\scriptstyle \pm \ 1.12}$ & 77.91 ${\scriptstyle \pm \ 0.20}$& 10.92 ${\scriptstyle \pm \ 0.08}$ & 81.41 ${\scriptstyle \pm \ 0.81}$& 10.78 ${\scriptstyle \pm \ 0.71}$ \\
\rowcolor{cyan!10}
Proposed TTI + Euc-constrained & 77.61 ${\scriptstyle \pm \ 1.03}$ & 10.75 ${\scriptstyle \pm \ 0.78}$ & 85.49 ${\scriptstyle \pm \ 0.34}$& 23.66 ${\scriptstyle \pm \ 2.54}$ & 76.88 ${\scriptstyle \pm \ 1.67}$ & 9.60 ${\scriptstyle \pm \ 0.69}$ & 79.11 ${\scriptstyle \pm \ 2.25}$& 10.31 ${\scriptstyle \pm \ 0.59}$\\
\rowcolor{blue!8}
Proposed TTI + Cos-constrained & 77.63 ${\scriptstyle \pm \ 2.14}$  & 9.64 ${\scriptstyle \pm \ 2.40}$ & 87.93 ${\scriptstyle \pm \ 0.62}$ & 18.42 ${\scriptstyle \pm \ 1.16}$ & 75.71 ${\scriptstyle \pm \ 3.21}$ & 9.28 ${\scriptstyle \pm \ 0.23}$& 78.70 ${\scriptstyle \pm \ 1.33}$ & 10.05 ${\scriptstyle \pm \ 0.46}$ \\
\rowcolor{red!8}
Proposed TTI + Neurotoxin & 79.10 ${\scriptstyle \pm \ 1.36}$ & 12.34 ${\scriptstyle \pm \ 1.23}$ & 88.14 ${\scriptstyle \pm \ 0.37}$ & 21.87 ${\scriptstyle \pm \ 1.86}$ & 76.92 ${\scriptstyle \pm \ 2.01}$ & 11.17 ${\scriptstyle \pm \ 0.41}$ & 79.37 ${\scriptstyle \pm \ 0.20}$ & 11.40 ${\scriptstyle \pm \ 0.36}$ \\
\hline
\end{tabular}
\begin{minipage}{0.9\textwidth}
\footnotesize
\textit{Note:} For a $C$-class classification task, the chance-level ASR is approximately $1/C$. Thus, the chance-level ASR is approximately 9.1\%, 25\%, and 10\% for RML2016.10A, AGNews, and CIFAR-10, respectively.
\end{minipage}
\end{table*}

\begin{figure*}[!t]
    \centering
    \vspace{2mm}
    \begin{subfigure}[t]{0.246\textwidth}
        \centering
        \includegraphics[width=\linewidth,trim=0 0 0 30,clip]{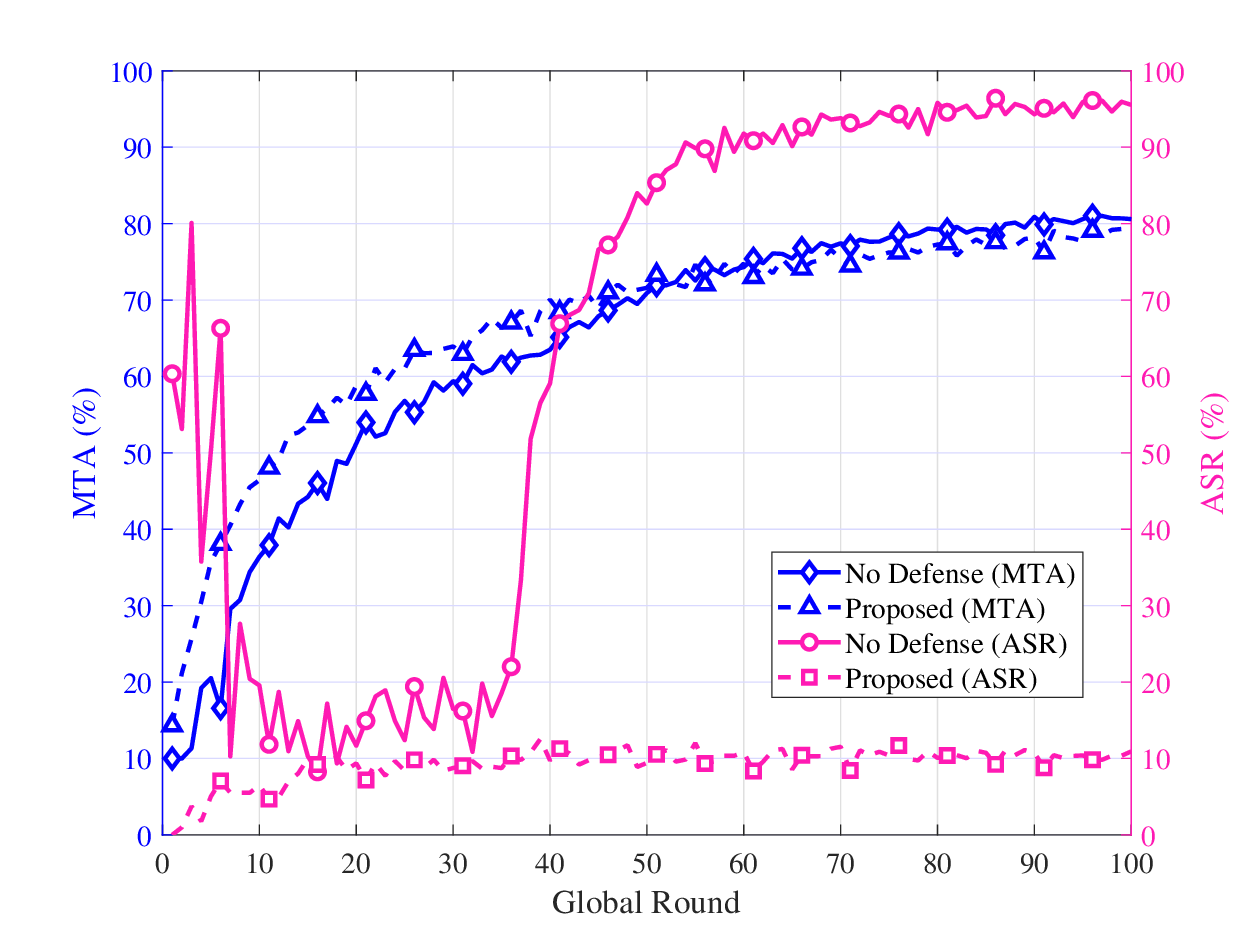}
        \caption{Bounded-scaling attack.}
        \label{fig:sub1}
    \end{subfigure}
    \hfill
    \begin{subfigure}[t]{0.246\textwidth}
        \centering
        \includegraphics[width=\linewidth,trim=0 0 0 30,clip]{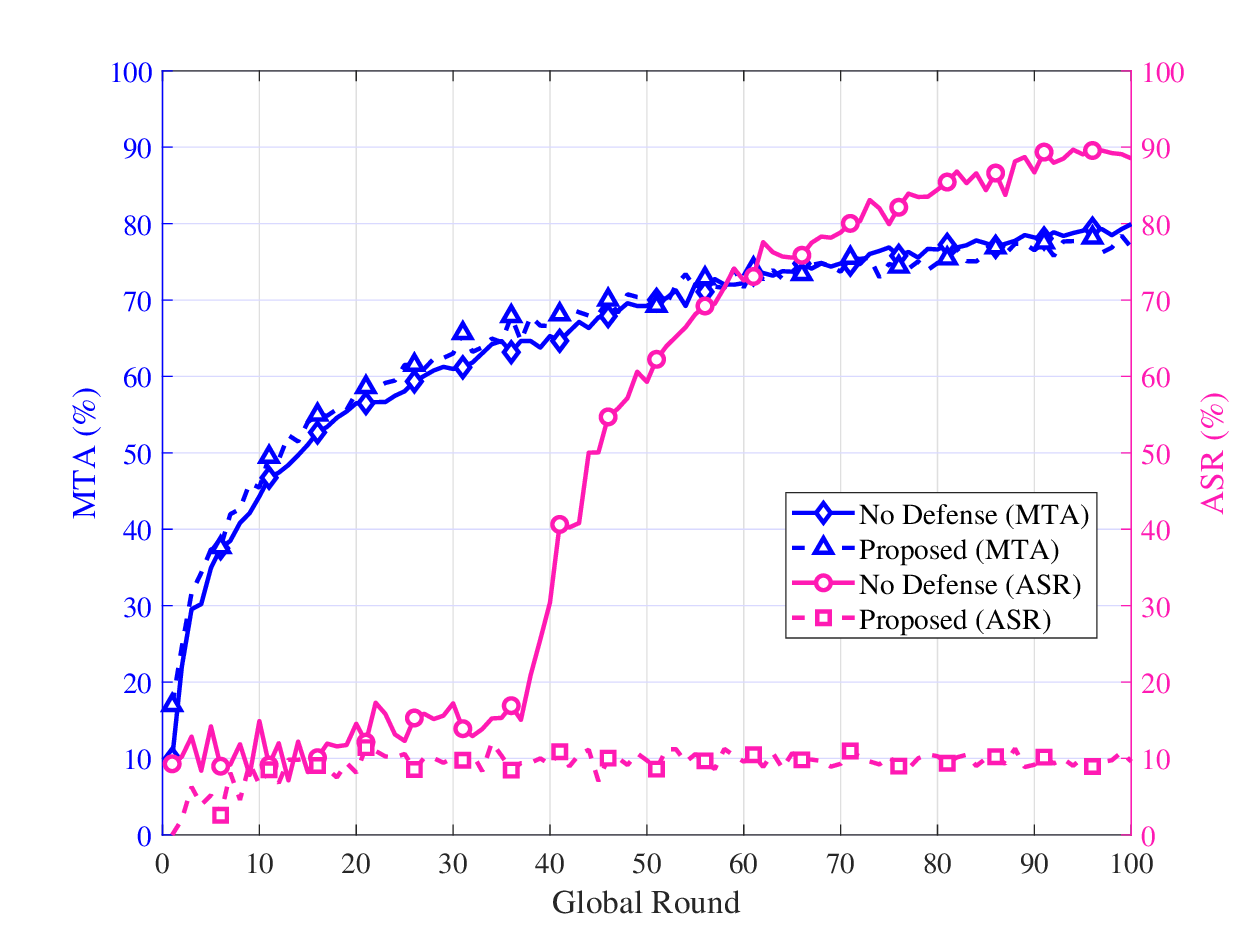}
        \caption{Euc-constrained attack.}
        \label{fig:sub2}
    \end{subfigure}
    \hfill
    \begin{subfigure}[t]{0.246\textwidth}
        \centering
        \includegraphics[width=\linewidth,trim=0 0 0 30,clip]{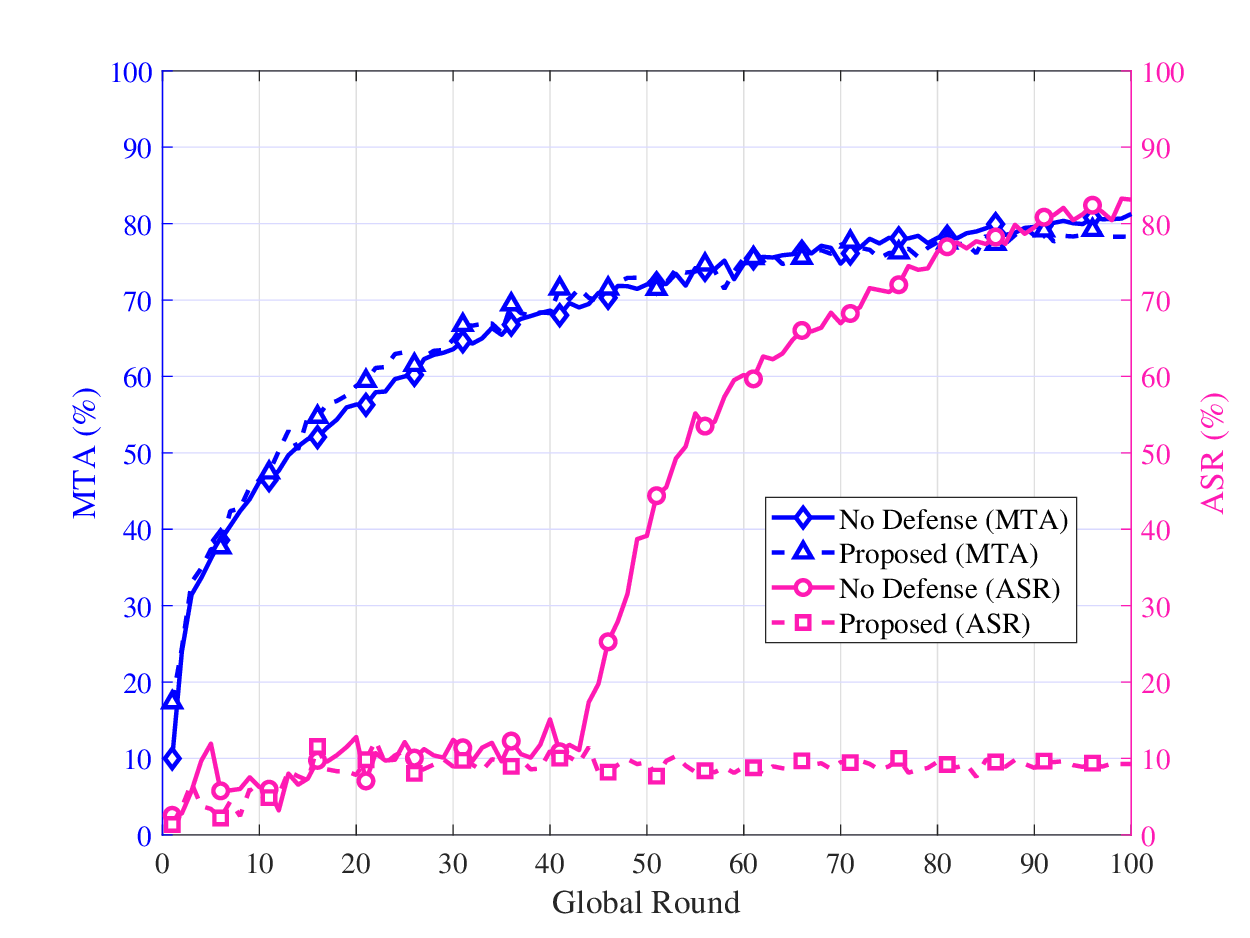}
        \caption{Cos-constrained attack.}
        \label{fig:sub3}
    \end{subfigure}
    \hfill
    \begin{subfigure}[t]{0.246\textwidth}
        \centering
        \includegraphics[width=\linewidth,trim=0 0 0 30,clip]{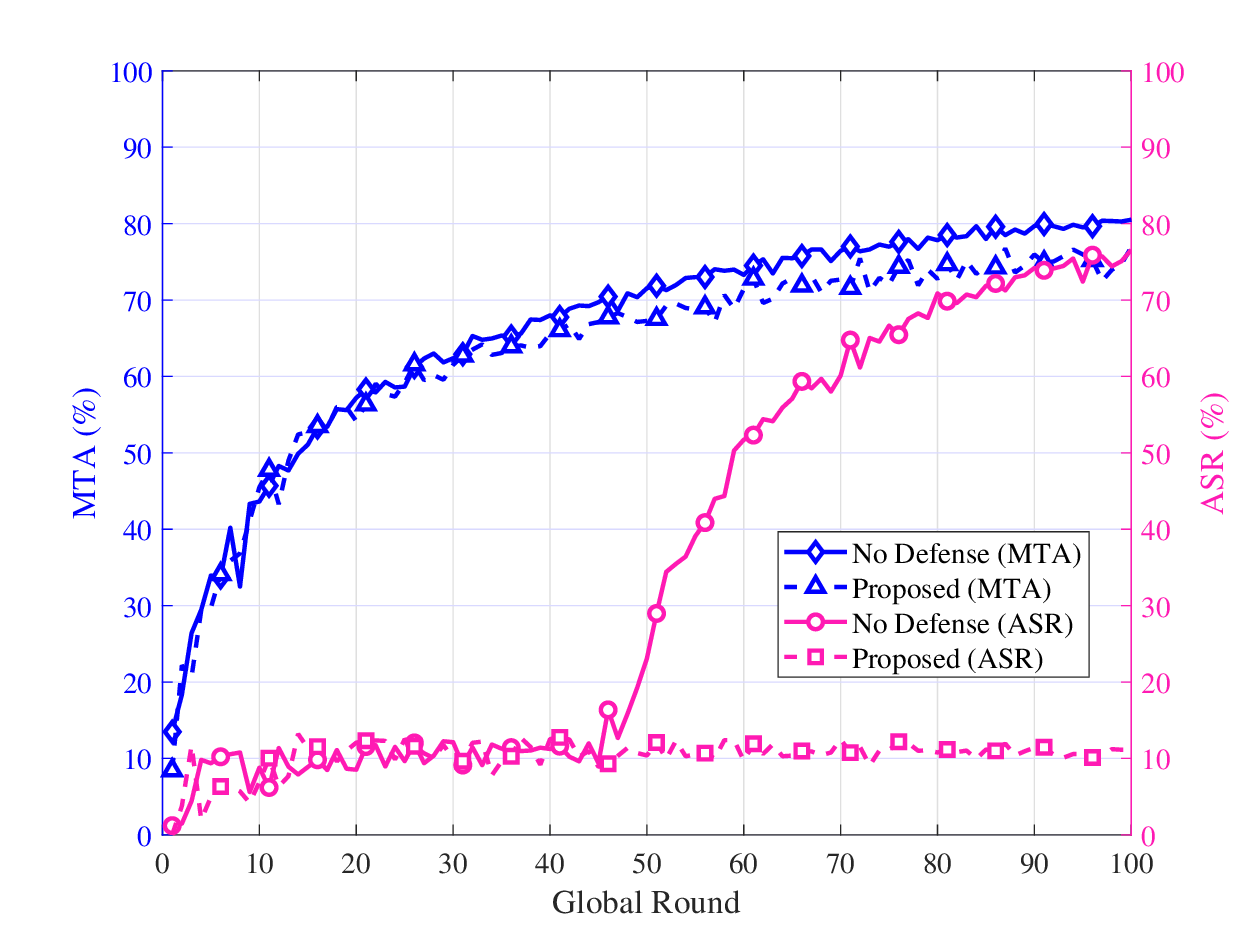}
        \caption{Neurotoxin attack.}
        \label{fig:sub4}
    \end{subfigure}
    \caption{Convergence behaviors of MTA and ASR for our method on the CIFAR-10 dataset with CNN training architecture.}
    \label{fig:convergence}
\end{figure*}

\section{Experimental Results}

\subsection{Experiment Setup}
\textbf{(1) Training Setup.}\\
\textit{(1.1) Datasets.} We use three representative datasets from different modalities, namely RML2016.10A, AG News, and CIFAR-10, corresponding to waveform signals, text, and image, respectively. \\
\textit{(1.2) Training model.} For RML2016.10A, we adopt a lightweight 1D CNN with four convolutional layers followed by a fully connected classifier. For AG News, we use a lightweight Text CNN consisting of an embedding layer, multiple 1D convolutional filters with different kernel sizes, and a final fully connected classifier. For CIFAR-10, we consider two representative image classification architectures, namely CNN and ResNet9, to verify that the proposed framework generalizes across different network structures.\\
\textit{(1.3) Modality-aware indicator instantiation.}~As discussed in Sec.~\ref{indicatordesign}, the Stage I trust score follows a unified weighted-indicator framework, while the specific indicator set is instantiated according to the data modality and model architecture. The corresponding dataset-specific indicator configurations are summarized in Table~\ref{tab:indicator_config}.\\
\textit{(1.4) Non-IID training data.} We focus on the non-IID setting. To simulate this practical scenario, we partition client data using a Dirichlet distribution $Dir(\alpha)$ over class labels. Unless otherwise specified, we set $\alpha=0.5$~\cite{alpha}.\\
\textit{(1.5) Client tiering strategy.} As discussed in Sec.~\ref{Tier_Strategy}, two client tiering strategies are available in the proposed framework. In simulations, we adopt the proportion-based strategy, as it guarantees a fixed number of clients in each tier regardless of the absolute trust score values. This design is more robust in dynamic multi-round OTA-FL training, where the trust score distribution may vary across communication rounds. The specific tiering ratio is (0.5, 0.3, 0.2).\\
\textit{(1.6) Malicious user setting.}
We consider a total of $K = 20$ clients in the OTA-FL system, among which $30\%$ are malicious users, i.e., $M=6$. These clients continuously launch backdoor attacks throughout the entire training process. This setting represents a persistent adversarial scenario where attackers consistently participate in the aggregation, posing a sustained threat to the global model.\\
\textbf{(2) Backdoor Attack Model.} In this paper, we consider the following backdoor attack models:
\begin{itemize}[leftmargin=*, itemsep=1pt, topsep=1pt, parsep=0pt]
    \item \textbf{Bounded-scaling attack} is a classical model poisoning strategy in which malicious clients amplify the magnitude of their local model updates before transmission, while constraining the resulting update norm within a bounded benign-looking range, so that the injected backdoor effect can still exert strong influence on the global aggregation~\cite{bagdasaryan2020backdoor}.
    
    \item \textbf{ Euclidean-constrained attack} is a stealthier variant that constrains the malicious update within a limited Euclidean distance from a benign-looking update, thereby reducing its detectability while preserving attack effectiveness \cite{bagdasaryan2020backdoor}.
    
\item \textbf{Cosine-constrained attack} improves stealthiness by forcing the malicious update to remain aligned with benign updates (i.e., high cosine similarity), while restricting its $l_2$ norm within a bounded range. In this way, the attack mimics benign drift under non-IID settings without causing gradient explosion \cite{bagdasaryan2020backdoor}.

\item \textbf{ Neurotoxin attack} is an advanced sparse backdoor attack that selectively perturbs only a small subset of model parameters, typically those that are less frequently updated, so as to implant the backdoor while minimizing its footprint in the overall model update \cite{zhang2022neurotoxin}.
\end{itemize}
These attack models cover a range of adversarial behaviors from aggressive amplification to well-crafted stealthy poisoning, providing comprehensive scenarios for evaluating the proposed framework.

\textbf{(3) Performance Indicator.} We consider the following two performance indicators \cite{bagdasaryan2020backdoor, Lee}:
\begin{itemize}[leftmargin=*, itemsep=1pt, topsep=1pt, parsep=0pt]
    \item \textbf{Main Task Accuracy (MTA)} indicates the accuracy of a model in its main task. The overall objective of all participants is to maximize MTA. The defense should not cause a significant drop in MTA and the attackers seek to be stealthy and maintain MTA.
    \item \textbf{Attack Success Rate (ASR)} measures the effectiveness of the backdoor task, i.e., the fraction of trigger-embedded inputs that are classified into the attacker-specified target label. A higher ASR means that the poisoned global model is more likely to map inputs containing the trigger to the designated malicious label. The attackers aim to maximize this metric, while an effective defense seeks to keep it low. 
\end{itemize}
\enlargethispage{\baselineskip}

\subsection{Effectiveness Validation of Proposed Scheme}
 Table \ref{tab:exp1_effectiveness} reports the MTA and ASR of the proposed framework under different attack models and datasets. It can be observed that, without defense, all considered attack strategies achieve high ASR while maintaining competitive MTA, demonstrating their strong effectiveness and stealthiness in OTA-FL systems. In contrast, the proposed TTI framework consistently reduces the ASR across all datasets and attack types. For example, the ASR is reduced from over 95\% to around 10\% on CIFAR-10 under bounded-scaling attack. Notably, for CIFAR-10, which is a 10-class classification task, an ASR around 10\% is already close to the chance-level prediction probability, suggesting that the trigger no longer provides a meaningful attack advantage. Meanwhile, the MTA remains competitive and only experiences marginal degradation compared with the no-attack baseline. These results validate that the proposed framework can effectively mitigate backdoor attacks while preserving the utility of the main learning task.

Figure~\ref{fig:convergence} shows the convergence behaviors of MTA and ASR on the CIFAR-10 dataset with CNN architecture under different backdoor attack models. We can observe that without defense, the ASR gradually increases during training and eventually stabilizes at a high level, indicating that the backdoor effect accumulates over global rounds and is successfully implanted into the global model. Meanwhile, the MTA continues to improve and converges normally, highlighting the stealthiness of these well-crafted backdoor attacks. \enlargethispage{\baselineskip}
By contrast, under the proposed TTI framework, the ASR remains consistently low across all attack models, while the MTA still converges stably to a level comparable to that of the no attack case. 
These results demonstrate that the proposed framework can effectively suppress malicious influence without sacrificing the convergence performance of the MTA.
These results in Table~\ref{tab:exp1_effectiveness} and Figure~\ref{fig:convergence} jointly verify that the proposed framework satisfies the three design objectives in Sec.~\ref{sec:framework}, namely effectiveness, utility, and generalizability. Specifically, TTI consistently suppresses the ASR across different attack models and datasets while maintaining competitive MTA, demonstrating strong robustness without sacrificing learning performance. Moreover, its consistent gains under diverse datasets, model architectures, and attack strategies further confirm its strong generalizability.

\begin{table*}[!t]
\centering
\caption{MTA  and ASR  comparisons on CIFAR-10 using CNN architecture under different defense schemes. We see that our method obtains the best combination of MTA and ASR across attack types.}
\label{tab:attack_vs_defense}
\setlength{\tabcolsep}{4pt}
\renewcommand{\arraystretch}{1.12}

\begin{tabular}{l|cc|cc|cc|cc}
\toprule

\multicolumn{1}{c|}{}
& \multicolumn{2}{>{\columncolor{gray!12}}c|}{\textbf{Bounded-Scaling}} 
& \multicolumn{2}{>{\columncolor{gray!12}}c|}{\textbf{Euc-constrained}} 
& \multicolumn{2}{>{\columncolor{gray!12}}c|}{\textbf{Cos-constrained}} 
& \multicolumn{2}{>{\columncolor{gray!12}}c}{\textbf{Neurotoxin}} \\

\multicolumn{1}{c|}{}
& \cellcolor{gray!12}\textbf{MTA ($\uparrow$)} & \cellcolor{gray!12}\textbf{ASR ($\downarrow$)}
& \cellcolor{gray!12}\textbf{MTA ($\uparrow$)} & \cellcolor{gray!12}\textbf{ASR ($\downarrow$)}
& \cellcolor{gray!12}\textbf{MTA ($\uparrow$)} & \cellcolor{gray!12}\textbf{ASR ($\downarrow$)}
& \cellcolor{gray!12}\textbf{MTA ($\uparrow$)} & \cellcolor{gray!12}\textbf{ASR ($\downarrow$)} \\
\midrule

\cellcolor{gray!12}\textbf{Proposed TTI}
& 77.91 ${\scriptstyle \pm \ 0.20}$  & 10.92 ${\scriptstyle \pm \ 0.08}$
& 76.88 ${\scriptstyle \pm \ 1.67}$ & 9.60 ${\scriptstyle \pm \ 0.69}$
& 75.71 ${\scriptstyle \pm \ 1.21}$ & 9.28 ${\scriptstyle \pm \ 0.23}$
&76.92 ${\scriptstyle \pm \ 2.01}$ & 11.17 ${\scriptstyle \pm \ 0.41}$ \\

\cellcolor{gray!12}TDA Tiering 
& 76.90 ${\scriptstyle \pm \ 1.06}$ & 11.15 ${\scriptstyle \pm \ 0.12}$
& 78.54 ${\scriptstyle \pm \ 0.67}$ & 10.78 ${\scriptstyle \pm \ 0.50}$
& \cellcolor{red!12}80.92 ${\scriptstyle \pm \ 0.63}$ &\cellcolor{red!12} 62.78 ${\scriptstyle \pm \ 2.44}$
&78.51 ${\scriptstyle \pm \ 0.86}$ & 16.09  ${\scriptstyle \pm \ 1.57}$\\

\cellcolor{gray!12}$\ell_2$ Norm Tiering
& \cellcolor{red!12}77.57 ${\scriptstyle \pm \ 2.01}$ & \cellcolor{red!12} 90.25 ${\scriptstyle \pm \ 0.56}$
& \cellcolor{red!12}73.65 ${\scriptstyle \pm \ 0.83}$ &  \cellcolor{red!12}77.41 ${\scriptstyle \pm \ 0.92}$
&\cellcolor{red!12}76.15 ${\scriptstyle \pm \ 0.93}$ & \cellcolor{red!12}46.43 ${\scriptstyle \pm \ 1.58}$
& 77.84 ${\scriptstyle \pm \ 1.12}$ & 10.94 ${\scriptstyle \pm \ 0.23}$ \\

\cellcolor{gray!12}Spikiness Tiering
& 75.79 ${\scriptstyle \pm \ 0.73}$ & 10.83 ${\scriptstyle \pm \ 0.31}$
& 76.62 ${\scriptstyle \pm \ 1.34}$ & 9.73 ${\scriptstyle \pm \ 0.65}$
& 75.86 ${\scriptstyle \pm \ 2.53}$ & 10.94 ${\scriptstyle \pm \ 0.33}$
&\cellcolor{red!12}72.68 ${\scriptstyle \pm \ 0.75}$ & \cellcolor{red!12}54.04 ${\scriptstyle \pm \ 0.21}$ \\


\cellcolor{gray!12}Model-wise Inspection
& 77.77 ${\scriptstyle \pm \ 0.79}$ & 11.58 ${\scriptstyle \pm \ 0.94}$
& 78.49 ${\scriptstyle \pm \ 0.19}$ & 10.28 ${\scriptstyle \pm \ 0.13}$
& 78.45 ${\scriptstyle \pm \ 0.24}$ & 11.61 ${\scriptstyle \pm \ 0.30}$
& \cellcolor{red!12}74.65 ${\scriptstyle \pm \ 0.72}$ & \cellcolor{red!12}62.10 ${\scriptstyle \pm \ 1.43}$ \\

\cellcolor{gray!12}BEV \cite{BEV_SGD}
&\cellcolor{red!12}79.04 ${\scriptstyle \pm \ 0.40}$& \cellcolor{red!12}89.50 ${\scriptstyle \pm \ 0.23}$
& \cellcolor{red!12}72.68 ${\scriptstyle \pm \ 1.61}$ & \cellcolor{red!12}54.49 ${\scriptstyle \pm \ 1.86}$
& \cellcolor{red!12}79.78 ${\scriptstyle \pm \ 0.87}$ & \cellcolor{red!12}45.24 ${\scriptstyle \pm \ 2.72}$
&\cellcolor{red!12}74.86 ${\scriptstyle \pm \ 1.23}$& \cellcolor{red!12}31.85 ${\scriptstyle \pm \ 0.61}$ \\

\cellcolor{gray!12}FedSAC \cite{group1}
& \cellcolor{red!12}77.03 ${\scriptstyle \pm \ 0.83}$& \cellcolor{red!12}96.90 ${\scriptstyle \pm \ 0.25}$
&\cellcolor{red!12} 77.75 ${\scriptstyle \pm \ 1.06}$ & \cellcolor{red!12}87.32 ${\scriptstyle \pm \ 0.67}$
& \cellcolor{red!12}80.70 ${\scriptstyle \pm \ 0.85}$ &\cellcolor{red!12} 80.3 ${\scriptstyle \pm \ 0.49}$
&\cellcolor{red!12} 79.88 ${\scriptstyle \pm \ 0.34}$& \cellcolor{red!12}76.67 ${\scriptstyle \pm \ 0.72}$ \\

\bottomrule
\end{tabular}

\begin{minipage}{0.8\textwidth}
\footnotesize
\textit{Note:} For a $C$-class classification task, the chance-level ASR is approximately $1/C$. Thus, the chance-level ASR is approximately 10\% for CIFAR-10.
\end{minipage}
\end{table*}

\begin{table*}[t]
\centering
\caption{Overhead comparison of the proposed TTI framework and representative defense baselines.}
\label{tab:overhead_compare}
\small
\renewcommand{\arraystretch}{1.08}
\setlength{\tabcolsep}{4pt}
\begin{tabular}{C{1.55cm}|C{6.1cm}|C{2.3cm}|C{2.1cm}|C{4.2cm}}
\hline
\textbf{Method} & \textbf{Main Extra Overhead} & \textbf{Affected Scope}& \textbf{Runtime/Round} & \textbf{Key Efficiency Implication} \\ 
\hline
Proposed TTI 
& Suspicious model upload and PS-side inspection 
& Suspicious tier only 
& 20.85\,s
&  Selective overhead in suspicious set \\
\hline
BEV \cite{BEV_SGD} 
& Tier-dependent weighted aggregation 
& System-wide
& 18.53\,s
& Weighting affects overall aggregation \\
\hline
FedSAC \cite{group1} 
& Client grouping, group-level aggregation, and scoring 
& System-wide
& 17.11\,s
& Repeated grouping and scoring \\ 
\hline
\end{tabular}
\begin{minipage}{\textwidth}
\footnotesize
\textit{Note:} Runtime/Round is measured as the average runtime per communication round over 100 rounds on CIFAR-10 with the CNN architecture under the Cos-constrained attack.
\end{minipage}
\end{table*}

\subsection{Comparison with Existing Methods}
In this subsection, we focus on CIFAR-10 dataset with CNN training architecture. And
we consider the following defense schemes to verify the effectiveness of the proposed framework:
\begin{itemize}[leftmargin=*, itemsep=1pt, topsep=1pt, parsep=0pt]
    \item \textbf{TDA Tiering, $\ell_2$ Norm Tiering, and Spikiness Tiering}: Ablation variants of the proposed TTI framework, where Stage I uses only one indicator for client tiering instead of the composite multi-indicator trust score. The trusted/suspicious/malicious partition and the subsequent Stage II inspection for suspicious clients remain unchanged.

    \item \textbf{Model-wise Inspection}: We replace the layer-wise PS-side inspection with a model-wise inspection. Specifically, Stage I remains unchanged, and in Stage II the same suspicious-client inspection procedure is retained, except that the proposed layer-wise inspection is replaced by a model-wise inspection.

    \item \textbf{Best-effort voting (BEV) based aggregation}: We consider an OTA-FL robust aggregation baseline based on best-effort voting~\cite{BEV_SGD}. After client tiering, trusted, suspicious, and denied clients are assigned aggregation weights of 1, 0.1, and 0, respectively, so that malicious influence is mitigated through weighted aggregation. The final global update is obtained by directly aggregating the weighted local updates without any additional PS-side inspection.\enlargethispage{\baselineskip}
       
    \item \textbf{Federated Learning with Secure Adaptive clustering (FedSAC):} a representative OTA-FL robust aggregation scheme based on reputation-guided grouping \cite{group1}. In each communication round, clients first compute their local updates and are then partitioned into 5 groups, where random grouping is used in the warm-up stage and reputation-guided grouping is adopted afterwards. The updates within each group are aggregated through OTA to form a group-level update. The PS then scores all group-level updates with respect to a clean reference update and retains only the top-ranked 3 groups for global aggregation.  
\end{itemize}

Table~\ref{tab:attack_vs_defense} compares the proposed TTI framework with the baselines described above. Specifically, the comparisons reveal three main observations: \\
\textit{(6.3.1) Comparison with single-indicator tiering.} The baselines that rely on only one indicator for client tiering exhibit clear weaknesses when the attack is designed to evade the corresponding metric. In particular, $\ell_2$ norm tiering performs poorly under bounded-scaling attack, Euc-constrained attack, and Cos-constrained attack. TDA tiering degrades significantly under Cosine-constrained attacks, and spikiness tiering becomes vulnerable under Neurotoxin attacks. This confirms that single-indicator designs can be easily bypassed by adaptive attackers, which justifies the proposed multi-indicator trust scoring mechanism. \\
\textit{(6.3.2) Comparison with model-wise inspection.} Replacing the proposed layer-wise PS-side inspection with whole-model inspection also degrades robustness, especially under the Neurotoxin attack. Since Neurotoxin perturbs only a small subset of less important parameters, its malicious effect is less visible at the whole-model level but more detectable through layer-wise structural abnormalities. This validates the advantage of the proposed layer-wise inspection design. \\
\textit{(6.3.3) Comparison with existing OTA-FL baselines.} Compared with BEV and FedSAC, the proposed TTI framework achieves substantially lower ASR under all attack settings. BEV improves robustness through weighted aggregation, but does not explicitly identify and exclude malicious clients, allowing stealthy backdoor updates to remain in the aggregation process. FedSAC performs group-level scoring and filtering, but lacks fine-grained per-client inspection, making it difficult to isolate carefully crafted malicious updates once they are mixed with benign users. We can observe from these results that the proposed scheme consistently maintains low ASR with competitive MTA, demonstrating strong effectiveness against backdoor attacks.
\enlargethispage{\baselineskip}

To further clarify the robustness-efficiency tradeoff of the proposed framework, Table~\ref{tab:overhead_compare} compares the overhead characteristics and runtime of the proposed TTI, BEV, and FedSAC. In TTI, the additional overhead is incurred only by suspicious clients, since only this tier is forwarded for individual uploading and PS-side layer-wise inspection. Under the proportion-based tiering ratio $(0.5,0.3,0.2)$ with $K=20$, at most 6 clients, i.e., 30\% of the participants, require such additional handling in each round. Although TTI has a slightly higher runtime per round, this extra cost is primarily due to the layer-wise PS-side inspection that enables more effective detection of stealthy malicious updates. As a result, TTI achieves a much lower ASR than BEV and FedSAC. Therefore, the proposed framework attains a favorable tradeoff by introducing slight additional runtime for a pronounced gain in robustness.

\begin{figure}
    \centering
    \includegraphics[width=\linewidth,trim=10 45 40 50,clip]{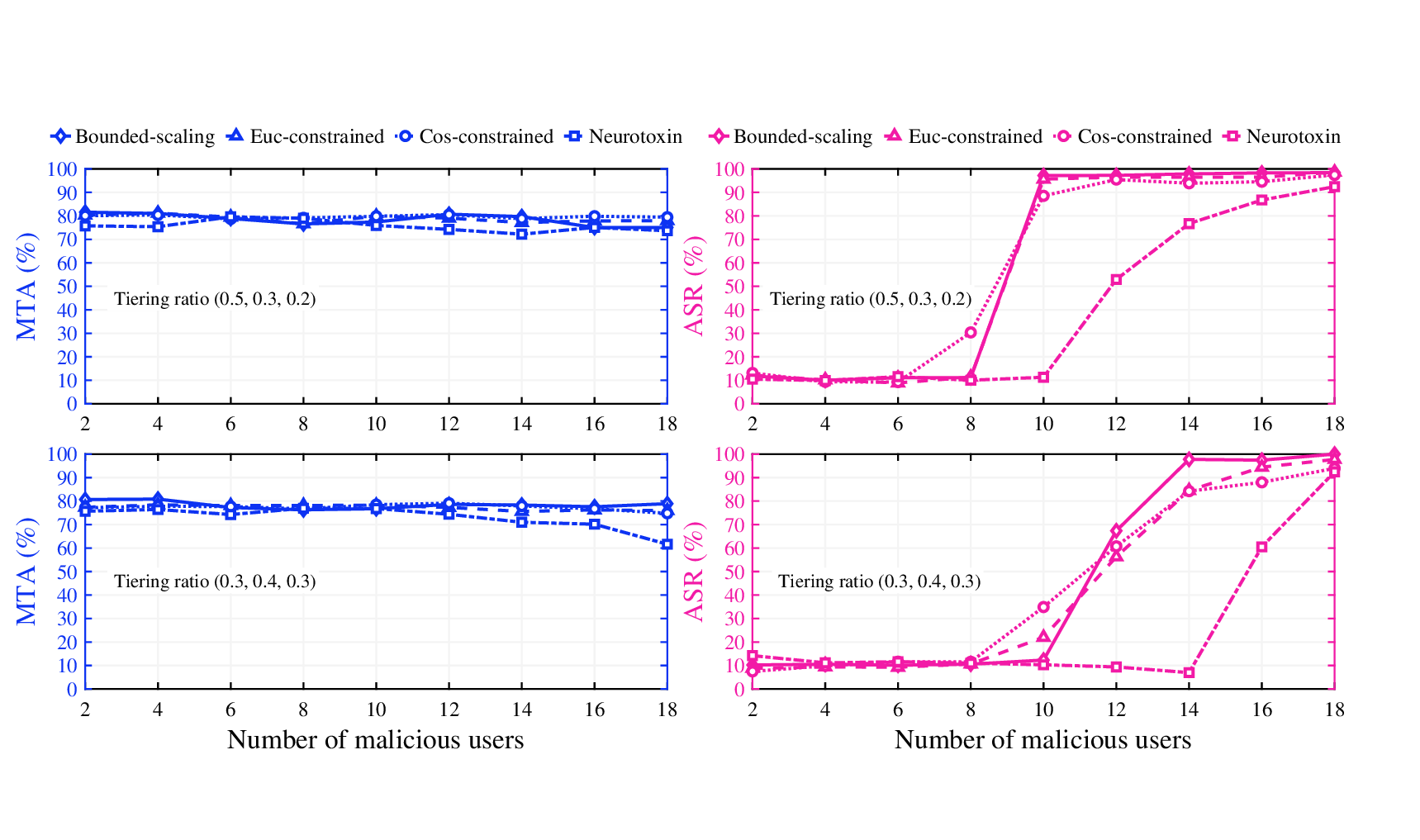}
    \caption{MTA and ASR for CIFAR-10 with the CNN model as the number of malicious users increases under different client tiering ratios in stage I. }
    \label{Fig:numberofmalicious}
\end{figure}
\subsection{Impact of the Number of Malicious Users}
 Figure~\ref{Fig:numberofmalicious} shows the MTA and ASR as the number of malicious users increases under two different client tiering ratios in stage I. We observe that the MTA remains comparatively stable across different attack models, indicating that the proposed framework maintains the utility of the main task even under stronger adversarial presence. For the ASR, a noticeable increase occurs only in the extreme regime, where the number of malicious users becomes overwhelmingly large. More importantly, near 100\% ASR is reached only when almost all participating clients are malicious. However, this setting is unrealistic for practical backdoor attacks, since if the vast majority of clients are already compromised, the global aggregation is naturally dominated by malicious updates, and thus a successful attack is expected regardless of the specific defense. In addition, comparing the two subfigures reveals that the tiering ratio has a clear influence on ASR. Specifically, the tiering ratio $(0.3, 0.4, 0.3)$ shown in the bottom row of Figure ~\ref{Fig:numberofmalicious} allocates a larger fraction of clients to the suspicious tier, so that more clients undergo PS-side layer-wise inspection before aggregation. Consequently, the resulting ASR is generally lower than that achieved by the ratio $(0.5, 0.3, 0.2)$ in the top row. This indicates that enlarging the set of clients subject to layer-wise inspection can further mitigate backdoor influence, at the expense of increased inspection overhead.

\section{Conclusion}
In this paper, we proposed a two-stage robust aggregation framework for defending against backdoor attacks in OTA-FL systems. By combining client-side trust-based tiering with PS-side layer-wise inspection, the proposed TTI framework effectively filters malicious updates while preserving the communication efficiency of OTA aggregation. Experimental results on multiple datasets, model architectures, and attack strategies show that TTI consistently reduces ASR while maintaining competitive MTA.

\FloatBarrier
\bibliographystyle{unsrt}
\bibliography{IntrusionRef}

\begin{thebibliography}{10}

\bibitem{vulnerability}
Yichen Wan, Youyang Qu, Wei Ni, Yong Xiang, Longxiang Gao, and Ekram Hossain.
\newblock Data and model poisoning backdoor attacks on wireless federated
  learning, and the defense mechanisms: {A} comprehensive survey.
\newblock {\em IEEE Communications Surveys \& Tutorials}, 26(3):1861--1897,
  2024.

\bibitem{ma2025}
Xiaoyan Ma, Shahryar Zehtabi, Taejoon Kim, and Christopher~G. Brinton.
\newblock Error analysis for over-the-air federated learning under misaligned
  and time-varying channels.
\newblock In {\em IEEE Global Communications Conference (GLOBECOM)}, 2025.

\bibitem{wang2025mitigating}
Su~Wang, Rajeev Sahay, Adam Piaseczny, and Christopher~G Brinton.
\newblock Mitigating evasion attacks in federated learning based signal
  classifiers.
\newblock {\em IEEE Transactions on Network Science and Engineering}, 2025.

\bibitem{Lee}
Seohyun Lee, Wenzhi Fang, Anindya Bijoy~Das, Seyyedali Hosseinalipour, David~J.
  Love, and Christopher~G. Brinton.
\newblock Cooperative decentralized backdoor attacks on vertical federated
  learning.
\newblock {\em IEEE Transactions on Networking}, 34:2004--2019, 2026.

\bibitem{bagdasaryan2020backdoor}
Eugene Bagdasaryan, Andreas Veit, Yiqing Hua, Deborah Estrin, and Vitaly
  Shmatikov.
\newblock How to backdoor federated learning.
\newblock In {\em International conference on artificial intelligence and
  statistics}, pages 2938--2948. PMLR, 2020.

\bibitem{no_indivudual}
Xiaowen Cao, Zhonghao Lyu, Guangxu Zhu, Jie Xu, Lexi Xu, and Shuguang Cui.
\newblock An overview on over-the-air federated edge learning.
\newblock {\em IEEE Wireless Communications}, 31(3):202--210, 2024.

\bibitem{gruoping}
Shuyan Hu, Xin Yuan, Wei Ni, Xin Wang, Ekram Hossain, and H.~Vincent~Poor.
\newblock {OFDMA-F²L: Federated learning with flexible aggregation over an
  OFDMA air interface}.
\newblock {\em IEEE Transactions on Wireless Communications}, 23(7):6793--6807,
  2024.

\bibitem{resource_block}
Zhanwei Wang, Kaibin Huang, and Yonina~C. Eldar.
\newblock Spectrum breathing: {P}rotecting over-the-air federated learning
  against interference.
\newblock {\em IEEE Transactions on Wireless Communications},
  23(8):10058--10071, 2024.

\bibitem{hierarchical}
Jiacheng Yao, Wei Shi, Wei Xu, Zhaohui Yang, A.~Lee Swindlehurst, and Dusit
  Niyato.
\newblock Byzantine-resilient over-the-air federated learning under zero-trust
  architecture.
\newblock {\em IEEE Journal on Selected Areas in Communications},
  43(6):1954--1969, 2025.

\bibitem{Krum}
Peva Blanchard, El~Mahdi El~Mhamdi, Rachid Guerraoui, and Julien Stainer.
\newblock Machine learning with adversaries: {B}yzantine tolerant gradient
  descent.
\newblock In {\em Advances in Neural Information Processing Systems}, 2017.

\bibitem{FoolsGold}
Clement Fung, Chris J.~M. Yoon, and Ivan Beschastnikh.
\newblock The limitations of federated learning in sybil settings.
\newblock In {\em 23rd International Symposium on Research in Attacks,
  Intrusions and Defenses (RAID 2020)}, pages 301--316, San Sebastian, October
  2020. USENIX Association.

\bibitem{FLTrust}
Xiaoyu Cao, Minghong Fang, Jia Liu, and Neil Gong.
\newblock {FLTrust: Byzantine-robust federated learning via trust
  bootstrapping}.
\newblock In {\em ISOC Network and Distributed System Security Symposium
  (NDSS)}, 01 2021.

\bibitem{FLDetector}
Zaixi Zhang, Xiaoyu Cao, Jinyuan Jia, and Neil~Zhenqiang Gong.
\newblock {FLdetector: Defending federated learning against model poisoning
  attacks via detecting malicious clients}.
\newblock In {\em Proceedings of the 28th ACM SIGKDD conference on knowledge
  discovery and data mining}, pages 2545--2555, 2022.

\bibitem{Flame}
Thien~Duc Nguyen, Phillip Rieger, Huili Chen, Hossein Yalame, Helen
  M{\"o}llering, Hossein Fereidooni, Samuel Marchal, Markus Miettinen, Azalia
  Mirhoseini, Shaza Zeitouni, Farinaz Koushanfar, Ahmad-Reza Sadeghi, and
  Thomas Schneider.
\newblock {FLAME}: Taming backdoors in federated learning.
\newblock In {\em 31st USENIX Security Symposium}, pages 1415--1432. USENIX
  Association, August 2022.

\bibitem{Shejwalkar2021Manipulating}
Virat Shejwalkar and Amir Houmansadr.
\newblock Manipulating the byzantine: Optimizing model poisoning attacks and
  defenses for federated learning.
\newblock In {\em Proc. Netw. Distrib. Syst. Secur. Symp. (NDSS)}, pages 1--18,
  2021.

\bibitem{blackbackdoor}
Thuy~Dung Nguyen, Tuan Nguyen, Phi Le~Nguyen, Hieu~H Pham, Khoa~D Doan, and
  Kok-Seng Wong.
\newblock Backdoor attacks and defenses in federated learning: {S}urvey,
  challenges and future research directions.
\newblock {\em Engineering Applications of Artificial Intelligence},
  127:107166, 2024.

\bibitem{xie2019dba}
Chulin Xie, Keli Huang, Pin-Yu Chen, and Bo~Li.
\newblock {DBA: Distributed backdoor attacks against federated learning}.
\newblock In {\em International conference on learning representations}, 2019.

\bibitem{wang2020attack}
Hongyi Wang, Kartik Sreenivasan, Shashank Rajput, Harit Vishwakarma, Saurabh
  Agarwal, Jy-yong Sohn, Kangwook Lee, and Dimitris Papailiopoulos.
\newblock Attack of the tails: {Y}es, you really can backdoor federated
  learning.
\newblock {\em Advances in neural information processing systems},
  33:16070--16084, 2020.

\bibitem{zhang2023a3fl}
Hangfan Zhang, Jinyuan Jia, Jinghui Chen, Lu~Lin, and Dinghao Wu.
\newblock {A3FL: Adversarially adaptive backdoor attacks to federated
  learning}.
\newblock {\em Advances in neural information processing systems},
  36:61213--61233, 2023.

\bibitem{li20233dfed}
Haoyang Li, Qingqing Ye, Haibo Hu, Jin Li, Leixia Wang, Chengfang Fang, and Jie
  Shi.
\newblock {3DFed: Adaptive and extensible framework for covert backdoor attack
  in federated learning}.
\newblock In {\em 2023 IEEE symposium on security and privacy (SP)}, pages
  1893--1907. IEEE, 2023.

\bibitem{zhang2022neurotoxin}
Zhengming Zhang, Ashwinee Panda, Linyue Song, Yaoqing Yang, Michael Mahoney,
  Prateek Mittal, Ramchandran Kannan, and Joseph Gonzalez.
\newblock {Neurotoxin: Durable backdoors in federated learning}.
\newblock In {\em International conference on machine learning}, pages
  26429--26446. PMLR, 2022.

\bibitem{BEV_SGD}
Xin Fan, Yue Wang, Yan Huo, and Zhi Tian.
\newblock {BEV-SGD: Best effort voting SGD against byzantine attacks for
  analog-aggregation-based federated learning over the air}.
\newblock {\em IEEE Internet of Things Journal}, 9(19):18946--18959, 2022.

\bibitem{group0}
Houssem Sifaou and Geoffrey~Ye Li.
\newblock Robust federated learning via over-the-air computation.
\newblock In {\em 2022 IEEE 32nd International Workshop on Machine Learning for
  Signal Processing (MLSP)}, pages 1--6, 2022.

\bibitem{group1}
Jiacheng Yao, Wei Shi, Wei Xu, Zhaohui Yang, A.~Lee Swindlehurst, and Dusit
  Niyato.
\newblock Byzantine-resilient over-the-air federated learning under zero-trust
  architecture.
\newblock {\em IEEE Journal on Selected Areas in Communications},
  43(6):1954--1969, 2025.

\bibitem{group2}
David Nordlund, Jialing Liao, and Zheng Chen.
\newblock Byzantine-resilient hierarchical federated learning with clustered
  over-the-air aggregation.
\newblock In {\em 2024 IEEE International Conference on Acoustics, Speech, and
  Signal Processing Workshops (ICASSPW)}, pages 715--719, 2024.

\bibitem{dummy0}
David Nordlund, Zheng Chen, and Erik~G. Larsson.
\newblock Detecting active attacks in over-the-air computation using dummy
  samples.
\newblock In {\em 2023 57th Asilomar Conference on Signals, Systems, and
  Computers}, pages 1691--1696, 2023.

\bibitem{dummy1}
Hang Zhou, Yi-Han Chiang, Caijuan Chen, Xiaoyan Wang, and Yusheng Ji.
\newblock Detecting model poisoning attacks via dummy symbol insertion for
  secure over-the-air federated learning.
\newblock In {\em 2025 IEEE 22nd Consumer Communications \& Networking
  Conference (CCNC)}, pages 1--6, 2025.

\bibitem{BC0}
Chuan Ma, Jun Li, Long Shi, Ming Ding, Taotao Wang, Zhu Han, and H.~Vincent
  Poor.
\newblock When federated learning meets blockchain: {A} new distributed
  learning paradigm.
\newblock {\em IEEE Computational Intelligence Magazine}, 17(3):26--33, 2022.

\bibitem{BC1}
Nanqing Dong, Zhipeng Wang, Jiahao Sun, Michael Kampffmeyer, William
  Knottenbelt, and Eric Xing.
\newblock Defending against poisoning attacks in federated learning with
  blockchain.
\newblock {\em IEEE Transactions on Artificial Intelligence}, 5(7):3743--3756,
  2024.

\bibitem{LWG}
Wael Issa, Nour Moustafa, Benjamin Turnbull, and Zahir Tari.
\newblock {LGP}: {L}ayerwise gradient purify for robust federated learning
  against poisoning attacks.
\newblock {\em IEEE Transactions on Dependable and Secure Computing},
  23(1):175--192, 2026.

\bibitem{alpha}
Shilong Wang, Jianchun Liu, Hongli Xu, Chenxia Tang, Qianpiao Ma, and Liusheng
  Huang.
\newblock Towards communication-efficient decentralized federated graph
  learning over {Non-IID} data.
\newblock {\em IEEE Transactions on Mobile Computing}, pages 1--17, 2025.

\end{thebibliography}
\end{document}